\documentclass[amssymb,aps,pra,twocolumn,footinbib,10pt,floatfix]{revtex4-1}

\usepackage{subfigure}
\usepackage{enumitem}
\usepackage[colorlinks=true,
			linkcolor=blue,
			urlcolor=blue,
			citecolor=blue]{hyperref}
\usepackage{graphicx}
\usepackage{bm}
\usepackage{mathtools}

\usepackage[overload]{empheq}

\renewcommand{\d}{\mathrm{d}}

\begin{document}

\title{Fermi arcs and surface criticality in dirty Dirac materials}

\author{Eric Brillaux}
\author{Andrei A. Fedorenko}%
\affiliation{\mbox{Univ Lyon, ENS de Lyon, Univ Claude Bernard, CNRS, Laboratoire de Physique, F-69342 Lyon, France}}
\date{\today}

\begin{abstract}
We study the effects of disorder on semi-infinite Weyl and Dirac semimetals where the presence of a boundary leads to the formation of either Fermi arcs/rays or Dirac surface states.
Using a local version of the self-consistent Born approximation, we calculate the profile of the local density of states and the surface group velocity. 
This allows us to explore the full phase diagram as a function of boundary conditions and disorder strength.  
While in all cases we recover the sharp criticality in the bulk, we 
unveil a critical behavior at the surface of Dirac semimetals, which is smoothed out by Fermi arcs in Weyl semimetals. 
\end{abstract}
\maketitle

\emph{\label{sec:intro} Introduction.} -- 
Three dimensional nodal semimetals are materials where several energy bands cross linearly at isolated points in the Brillouin zone: two bands in Weyl semimetals~\cite{Xu:2015a,Xu:2015b}, and four bands in Dirac semimetals~\cite{Liu:2014,Neupane:2014,Borisenko:2014}. They exhibit remarkable phenomena related to the relativistic nature of the low energy excitations and topological properties of band structure~\cite{Yang2011,Burkov2016,Armitage2018,Son2013}. 
In particular, the bulk-boundary correspondence leads in Weyl semimetals to topologically protected surface-localized states in the form of Fermi arcs that connect the surface projections of Weyl nodes with opposite chirality~\cite{Balents:2011,Wan:2011,Wang:2012}.
They have been observed using photoemission spectroscopy in inversion-symmetry-breaking crystals such as tantalum arsenide (TaAs)~\cite{Xu:2015a,Lv:2015} and niobium arsenide (NbAs)~\cite{Xu:2015b}, where their shape and topological properties agree beautifully with first-principle calculations~\cite{Hasan2017}.
In Dirac semimetals, scattering from the boundary can also produce propagating surface modes with energies near the bulk band crossing
which, however, are not topologically protected~\cite{Armitage2018}.  
Experimentally, Dirac semimetals host at least one pair of Dirac nodes (like in $\mathrm{Na}_3 \mathrm{Bi}$), so that the Fermi surface may consist of two arcs that bridge the two bulk nodes~\cite{Xu2015-2}.
The local properties of the emergent surface states are controlled by the boundary conditions,
which describe how the different degrees of freedom such as pseudospin and valley index mix upon quasiparticle reflection from the surface~\cite{ShtankoLevitov2018,hashimoto_boundary_2017,faraei_greens_2018}.

The presence of disorder such as lattice defects or impurities can strongly modify the behavior of clean materials, or even lead to quantum phase transitions such as Anderson localization~\cite{Abrahams:2010}. A new type of disorder-induced quantum phase transition was recently discovered in relativistic semimetals~\cite{Syzranov:2018}, wherein a strong enough disorder drives the semimetal towards a diffusive metal. Inside the bulk, the average density of states (DoS) at the nodal point plays the role of an order parameter, since it becomes nonzero above a critical disorder strength~\cite{Sbierski:2014, Sbierski:2016,Fradkin:1986,Roy:2016b,Goswami:2011,Hosur:2012,Ominato:2014,Chen:2015,Altland:2015:2016}. This bulk transition has been intensively studied using both numerical
simulations~\cite{Kobayashi:2014,Sbierski:2015,Liu:2016,Bera:2016,Fu:2017,Sbierski:2017}
and analytical methods~\cite{Syzranov:2015b,Roy:2014,Louvet:2016,Balog:2018,Louvet:2017,Sbierski:2019,Klier:2019}.
The effects of rare events have also been much debated~\cite{Holder:2017,Gurarie:2017,Nandkishore:2014,Pixley:2016,Pixley:2016c,Wilson:2017,Wilson:2018,Ziegler:2018,Buchhold:2018,Buchhold:2018-2}.

How disorder affects the surface states of relativistic semimetals is much less known.
Perturbative calculations show that the surface states in generic Dirac materials  are  protected from surface disorder due to 
a slow decay of the states from the surface~\cite{ShtankoLevitov2018}.
Numerical simulations also indicate that while Fermi arcs in Weyl semimetals are robust against weak bulk disorder;
they hybridize with nonperturbative bulk rare states as the strength of disorder gradually increases 
and completely dissolve into the emerging metallic bath at the bulk transition~\cite{Slager2017,wilson_surface_2018}. 

In this Letter we study the effect of weak disorder on the surface states produced by generic boundary conditions in a minimal model for both Weyl and Dirac semimetals. We develop a local version of the self-consistent Born approximation (SCBA)~\cite{Klier:2019}, which enables us to compute the DoS profile
and the surface group velocity for different disorder strengths. We then investigate the full phase diagram in the presence of a surface, and show that to some extent it is similar to the one for semi-infinite magnetic systems, which exhibit ordinary, surface and extraordinary phase transitions~\cite{lubensky_critical_1975, lubensky_critical_1975_a}.
In particular, 
we find that when the boundary of a nodal semimetal hosts single-cone Dirac surface states, it turns into a metallic state at a critical disorder strength which is lower than that for the bulk.
Upon further increasing disorder, the bulk, in turn, becomes metallic through the transition which is called extraordinary, since the bulk ordering 
takes place in the presence of the ordered surface.
Surface and extraordinary transition lines meet at the special point. 
In material realizations of Weyl and Dirac semimetals, the surface criticality is smoothed out by the presence of surface states that populate the Fermi arcs even in the clean sample.

\emph{\label{sec:model} Boundary conditions for a semi-infinite semimetal.}~--
The Nielsen-Ninomiya theorem constrains relativistic semimetals to host pairs of nodes with opposite Berry charges~\cite{Nielsen:1981}. A minimal low energy theory for such materials thus comprises two Weyl nodes of oppo\-site chiralities, separated in the Brillouin zone by a momentum $2\bm{b}$. Below a suitable cut-off momentum $\Lambda$, the quasiparticles are determined by the binodal Weyl Hamiltonian~\cite{AltlandBagrets:2016}
\begin{equation}\label{eq:bulk_H}
H_0 = i \tau_z \bm{\sigma} \! \cdot \! \bm{\partial} + \tau_0 \bm{\sigma} \! \cdot \! \bm{b},
\end{equation}
where $\bm{\partial}=(\partial_x,\partial_y,\partial_z)$ denotes the gradient operator. In Eq.~\eqref{eq:bulk_H}, we use the Pauli matrices $\bm{\sigma} = (\sigma_x,\sigma_y,\sigma_z)$ and $(\tau_x,\tau_y, \tau_z)$ for the pseudospin and valley -- or chiral -- degrees of freedom, respectively. The identity matrices are denoted as $\sigma_0$, $\tau_0$.
We can formally tune $\bm{b}$ to describe either a
pair of Weyl cones ($\bm{b}$ arbitrary) or a single Dirac cone ($\bm{b}=\bm{0}$).
Notice that to prevent the two Weyl nodes from hybridizing and opening a gap when $\bm{b}$ vanishes, additional space symmetries must be present.

In a semi-infinite material filling the $z\geq 0$ half-space, we must supplement the bulk Hamiltonian~\eqref{eq:bulk_H} with proper boundary conditions (BC) at the surface. Assuming a generic BC that can describe not only a free surface, but also
a surface that is covered by various chemical layers, we impose
\begin{equation}\label{eq:BC} 
M \psi|_{z=0^+} =  \psi|_{z=0^+},
\end{equation}
where the unitary hermitian matrix $M$ ensures the nullity of the transverse current : $\{M,\tau_z \sigma_z\}=0$~\cite{hashimoto_boundary_2017,
faraei_greens_2018}. Two classes
of matrices satisfy these criteria. They are both parametrized by a pair of angles~$(\theta_+,\theta_-)$ such that
\begin{subequations}\label{eq:euler}
\begin{empheq}[left={\empheqlbrace\,}]{align}
M_1 & = \tau_+ (\bm{\sigma} \cdot \bm{e_{+}}) + \tau_{-}(\bm{\sigma} \cdot \bm{e_{-}}),\label{eq:M1}\\
M_2  & = (\bm{\tau} \cdot \bm{e_{+}} ) \sigma_{+} + (\bm{\tau} \cdot \bm{e_{-}}) \sigma_{-}, \label{eq:M2}
\end{empheq}
\end{subequations}
where $\bm{e_{\pm}} = (\cos \theta_\pm ,\sin \theta_\pm,0)$ are two unitary vectors of the surface, and $\mu_\pm = (\mu_x \pm i\mu_y)/2$ with $\mu = \tau, \sigma$ are the chiral and pseudospin projectors.

\begin{figure}[t!]
\centering
\subfigure{\label{fig:FA_M1} \includegraphics[scale=1]{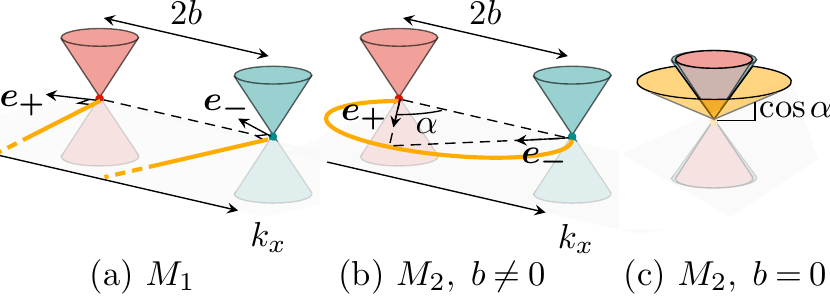}}
\subfigure{\label{fig:FA_M2}}
\subfigure{\label{fig:Semicone}}\\[-0.5cm]
\caption{Projection to the surface Brillouin zone $(k_x,k_y)$ of bulk Weyl cones with chiralities $+1$ (red) and $-1$ (blue). Surface states disperse along lines at the Fermi level (orange). (a) Fermi rays oriented by the vectors $\bm{e_\pm}$. (b) Fermi arc with curvature $\alpha$. (c) Surface cone with Fermi velocity $\cos \alpha$.}
\end{figure}

The BC~\eqref{eq:M1} describes Weyl fermions that retain the same chirality under scattering from the boundary. The emerging surface states at the Fermi level distribute along two independent Fermi rays pointing in the directions orthogonal to $\bm{e_{\chi}}$ stemming from the nodes of topological charges $\chi=\pm 1$ respectively, regardless of the distance between the nodes, as illustrated in Fig.~\ref{fig:FA_M1}. This BC breaks the $O(2)$ rotational symmetry in the ($x,y$) plane by imposing the rays' orientations, which must be determined by microscopic details of the boundary, and thus should be extremely sensitive to surface roughness~\cite{witten_three_2016}. It also mixes neighboring Landau levels in a background magnetic field, so that the Landau quantization is ill-defined~\cite{faraei_induced_2019}. Seemingly infinite Fermi rays extend to the full Brillouin zone and thus could terminate at another remote pair of Weyl nodes~\cite{hashimoto_boundary_2017}. 

The BC~\eqref{eq:M2}, on the contrary, mixes chirality of reflected quasiparticles, and thus depends qualitatively on the relative positions of the Weyl nodes. Without loss of generality, we align the nodes in the $x$ direction taking $\bm{b} = b\bm{e_x} + b_z \bm{e_z}$. (i) For a nonzero half-separation $b$ between the surface projections of the nodes, the surface states at the Fermi level disperse along a curved Fermi arc. Its parametric equation reads $\phi_+ - \phi_- + \theta_+ - \theta_- = 0$, where $\phi_{\chi}$ is the angle formed by the momentum measured from the node of chirality $\chi$ with the $x$ axis. This defines a circular arc with aperture angle $4\alpha$ and perimeter $L_{\rm FA} = 4b\alpha/\sin(2\alpha)$, as shown in Fig.~\ref{fig:FA_M2}, where the angle $\alpha = (\pi+\theta_+-\theta_-)/2 \in [0,\pi/2]$ reduces here to $\theta_+$ due to the alignement of the nodes along the $x$ axis. In experiments~\cite{xu_experimental_2015, lv_experimental_2015, inoue_quasiparticle_2016}, Fermi arcs are usually distorted because of higher-order corrections to the linear dispersion relation. They can join two Weyl nodes (our model) but also two Dirac nodes or surface-projected nodes with higher topological charges, in which case multiple arcs are attached to the pair. (ii) When $b=0$, the surface states are nontopological and form a single cone with Fermi velocity $v_0= \cos \alpha$ that extends in either the electron or hole side depending on the direction of the normal to the surface~\cite{ShtankoLevitov2018}. In our case, this corresponds to the positive energies, as shown in Fig.~\ref{fig:Semicone}.
This electron-hole surface asymmetry persists in the presence of a magnetic field or a gap
where the dispersion relation also depends on the extra parameter $\theta_{\tau} = (\theta_+ + \theta_-)/2$~\cite{ShtankoLevitov2018}.

\emph{\label{sec:disorder} Treatment of disorder.} -- 
Point-like impurities generate disorder that is insensitive to the chiral and pseudospin degrees of freedom. Assuming the density of impurities is uniform in the bulk, we model such defects by a random, scalar, gaussian potential $V(\bm{r})$ with zero average and short-range variance
$\gamma \delta(\bm{r})$. In an infinite sample, disorder induces a second-order transition towards a diffusive metal phase above a nonzero critical value~$\gamma^*$, though the critical point is probably avoided due to rare events~\cite{Holder:2017,Gurarie:2017,Nandkishore:2014,Pixley:2016,Pixley:2016c,Wilson:2017,Wilson:2018,Ziegler:2018,Buchhold:2018,Buchhold:2018-2}. The bulk average DoS $\bar{\rho}_{\rm b}$, which vanishes on the semimetal side, increases in a power-law fashion $\bar{\rho}_{\rm b} \sim (\gamma-\gamma^*)^{\beta}$ above the critical point~\cite{Sbierski:2014, Sbierski:2016,Fradkin:1986,Roy:2016b,Goswami:2011,Hosur:2012,Ominato:2014,Chen:2015,Altland:2015:2016}. As shown below, the spatially resolved DoS reveals this behavior for all BC, but also a new surface transition in a single Dirac cone.

\emph{\label{sec:LSCBA} Local self-consistent Born approximation.} -- 
The boundary breaks translational invariance along the perpendicular direction. Consequently, the retarded Green's function of the clean system $G_0(\epsilon,z,z')$
 depends not only on the distance $z-z'$ between the points but also on the absolute distance $z+z'$ to the surface. It satisfies the boundary condition $M G_0 (\epsilon,0,z')=G_0(\epsilon,0,z')$. Notice that we omit the explicit dependence on the momentum $k$ parallel to the boundary for the sake of brevity.  
Introducing the disorder-averaged Green's function $G(\epsilon,z,z')$, which satisfies the same boundary condition, we define the corresponding  self-energy  $\Sigma$ as
\begin{equation} \label{eq:Gdef}
[H_0 - (\epsilon  -\Sigma(\epsilon,z))\tau_0 \sigma_0]G(\epsilon,z,z') = \delta(z-z') \tau_0 \sigma_0,
\end{equation}
Within the SCBA and for point-like disorder  the self-energy $\Sigma$ is  momentum-independent in the bulk and proportional to the unit matrix~\cite{Klier:2019}.
In the absence of translational invariance it is a function of $z$ and satisfies the self-consistency equation   
\begin{equation}\label{eq:SCBA} 
\Sigma(\epsilon, z) = -\dfrac{\gamma}{4} \int_{|\bm{k}|<\Lambda} \dfrac{\d^2 k}{(2\pi)^2} \text{Tr} \left[G(\epsilon,z,z)\right].
\end{equation}
The solution to equations (\ref{eq:Gdef})-(\ref{eq:SCBA}), where the latter relates the self-energy at position $z$ to its values at all points, requires the inversion of a non-transitionally invariant Green's function. The problem is greatly simplified if one assumes that the spatial variations of the self-energy are small, that is to say if $\partial \Sigma/ \partial z \ll \Sigma^2$.
In this approximation we  replace $G(\epsilon,z,z)$ with $G_0(\epsilon-\Sigma(\epsilon,z),z,z)$  in Eq.~\eqref{eq:SCBA} so that the self-energy now satisfies a self-consistency equation whose both sides involve $\Sigma(\epsilon,z)$ at the \emph{same} position $z$.  We refer to this scheme as the local self-consistent Born approximation (LSCBA).

\begin{figure}[t!]
\subfigure{\label{fig:rho_M1}} \subfigure{\label{fig:rho_M2}}
\includegraphics[scale=0.65]{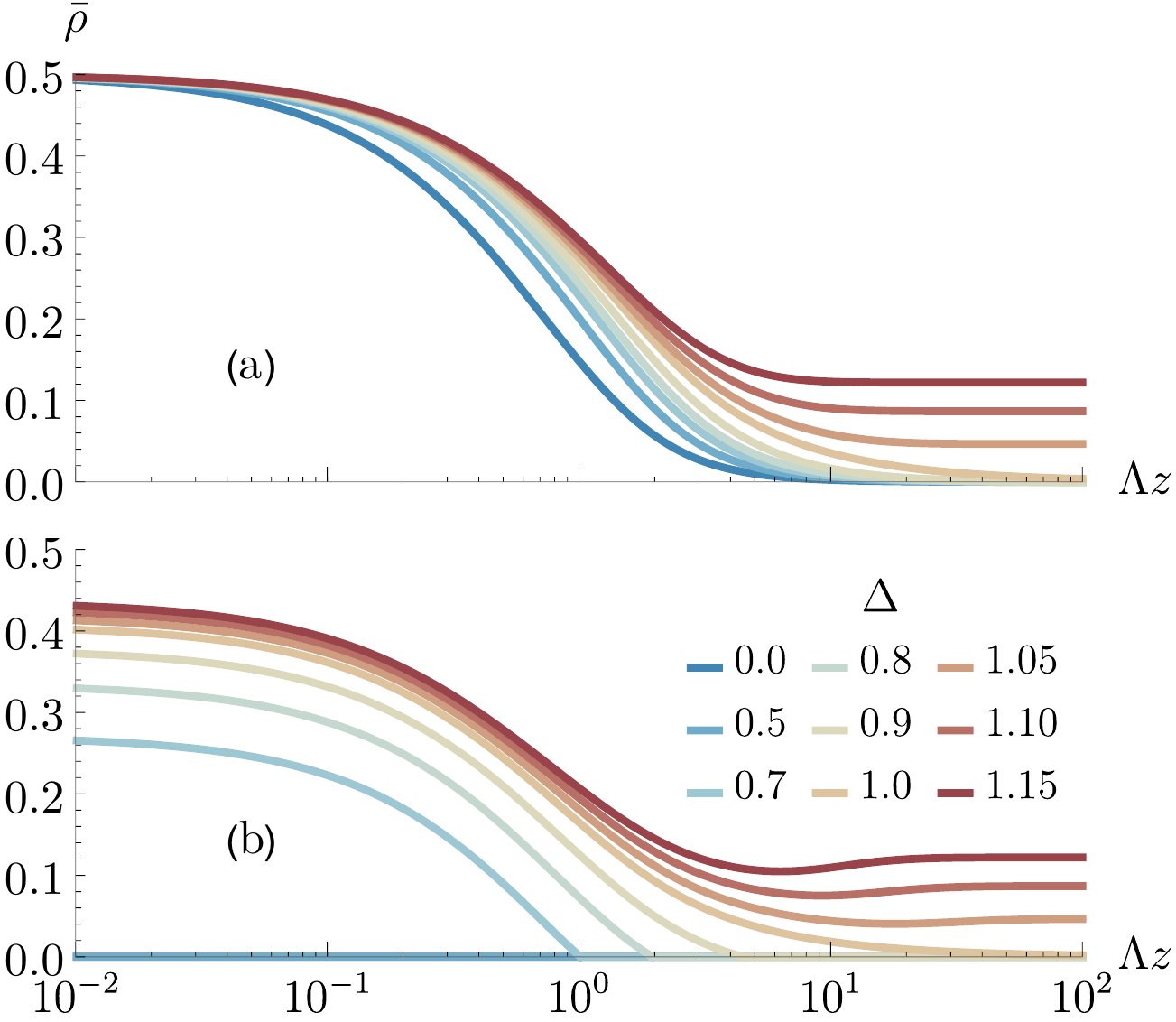}
\caption{LDoS profile for various disorder strengths $\Delta$. The bulk density increases as $\frac{1}{2}(1-\Delta^{-2})$ above the critical value $\Delta_{\rm b}^* = 1$, and vanishes below. (a) Fermi rays ($M_1$ BC). (b) Dirac surface states ($M_2$ BC) with $\alpha=\pi/4$.}
\end{figure}

In the presence of disorder, the surface band crossing arising from the $M_2$ BC is shifted from the bulk nodal energy  by  $\text{Re} \Sigma (\epsilon, z) $ similar to that found in~\cite{papaj_nodal_2019}. 
In this case it is natural instead of computing the density profile at a fixed chemical potential, e.g. at the bulk Fermi level, 
to calculate it at the energy of the local density minimum $\epsilon_{\rm F}$ given by $\epsilon_{\rm F}(z) =  \text{Re} \Sigma(\epsilon_{\rm F}(z),z)$.  This allows one to probe the  band crossing structure close to the surface as 
a function of the distance to the surface and  strength of disorder.
To that end we introduce the disorder-induced broadening $\Gamma>0$ as $\epsilon_{\rm F}(z)-\Sigma(\epsilon_{\rm F}(z),z) = i\Gamma(z)$, from which we extract the minimum of the average LDoS at pozition $z$, which in natural units reads
\begin{equation} \label{eq:DOS-profile}
\bar{\rho}(z) = -\dfrac{4\pi \text{Im}\Sigma(\epsilon_{\rm F}(z),z)}{\gamma \Lambda^2} = \dfrac{\Gamma(z)}{\Delta \Lambda},
\end{equation}
where $\Delta=\gamma \Lambda /4\pi$ is the dimensionless disorder strength.

\emph{\label{sec:DSS} Effect of disorder on Dirac surface states.} --
Let us focus
first on the case $b=0$.
Using the LSCBA~\eqref{eq:SCBA} we calculate the 
self-energy for both BC leading either to Fermi rays for $M_1$ or to Dirac surface states for $M_2$.
The corresponding LDoS profiles computed using Eq.~\eqref{eq:DOS-profile} are depicted in Fig.~\ref{fig:rho_M1} for Fermi rays, and in Fig.~\ref{fig:rho_M2} for Dirac surface states with $\alpha=\pi/4$. For both BC, we recover the bulk transition for infinite $z$ and at the critical disorder strength $\Delta_{\rm b}^* = 1$, above which the LDoS behaves~as $\bar{\rho}_{\rm b}
= \frac{1}{2} \left(1-\Delta^{-2}\right) \propto |\Delta-\Delta_{\rm b}^*|^\beta$
with $\beta=1$. Below this critical value, the LDoS vanishes, as in infinite systems within the SCBA~\cite{klier_weak_2019}. Exactly at criticality, the LDoS profile decreases algebraically as
$(\Lambda z)^{-1}$. 
    
Near the surface,
however, the LDoS behaves very differently depending on the BC. For Fermi rays, disorder does not impact the density close to the boundary, which remains always finite since the rays pass through the whole Brillouin zone (see Fig.~\ref{fig:FA_M1}).  The surface modes populating the rays propagate diffusively with a decreasing mean free path $l = 2/\Delta$ as we gradually increase the disorder strength, and dissolve into the metallic bulk above the semimetal-metal critical point.
    
On the contrary, the local density of Dirac surface states vanishes above a distance $\Lambda z = \xi$ from the boundary such that
\begin{equation}
\dfrac{1-e^{-2\xi}}{2\xi} = \dfrac{\Delta^{-1} - 1}{(\tan \alpha)^2}.
\end{equation}
Hence, Dirac surface states extend into the bulk over this penetration length $\xi$; they spread maximally at bulk criticality where it diverges like $\xi \sim |\Delta-\Delta_{\rm b}^*|^{-\nu}$ with $\nu=1$, which can be identified with the correlation
length exponent. 

Indeed, the values of the exponents $\beta$ and $\nu$ agree with the mean-field values of the critical exponents for the order parameter and correlation length  at the 3D semimetal-diffusive metal transition computed within the SCBA~\cite{klier_weak_2019}. In addition, $\xi$ vanishes at the disorder strength $\Delta_{\rm s}^*= (\cos \alpha)^2 < \Delta_{\rm b}^*$, below which the LDoS is zero everywhere. These cues signal that a surface transition takes place at $\Delta_{\rm s}^*$.

Before we consider  the full phase diagram, 
let us discuss the validity of the LSCBA. The local approximation is justified when the derivative of $\Gamma(z)$ is small with respect to $\Gamma(z)^2$, which is satisfied close to the surface and deep in the bulk. It breaks down 
at intermediate distance to the boundary $\Lambda z \sim 1$, 
where the solutions of the LSCBA for $\Lambda z=0$ and for $\Lambda z\gg 1$ match smoothly. 
Notice that the exact vanishing of the density for $\Lambda z>\xi$, 
is an artifact of this local approximation. 

\begin{figure}[t!]
\centering
\includegraphics[scale=0.65]{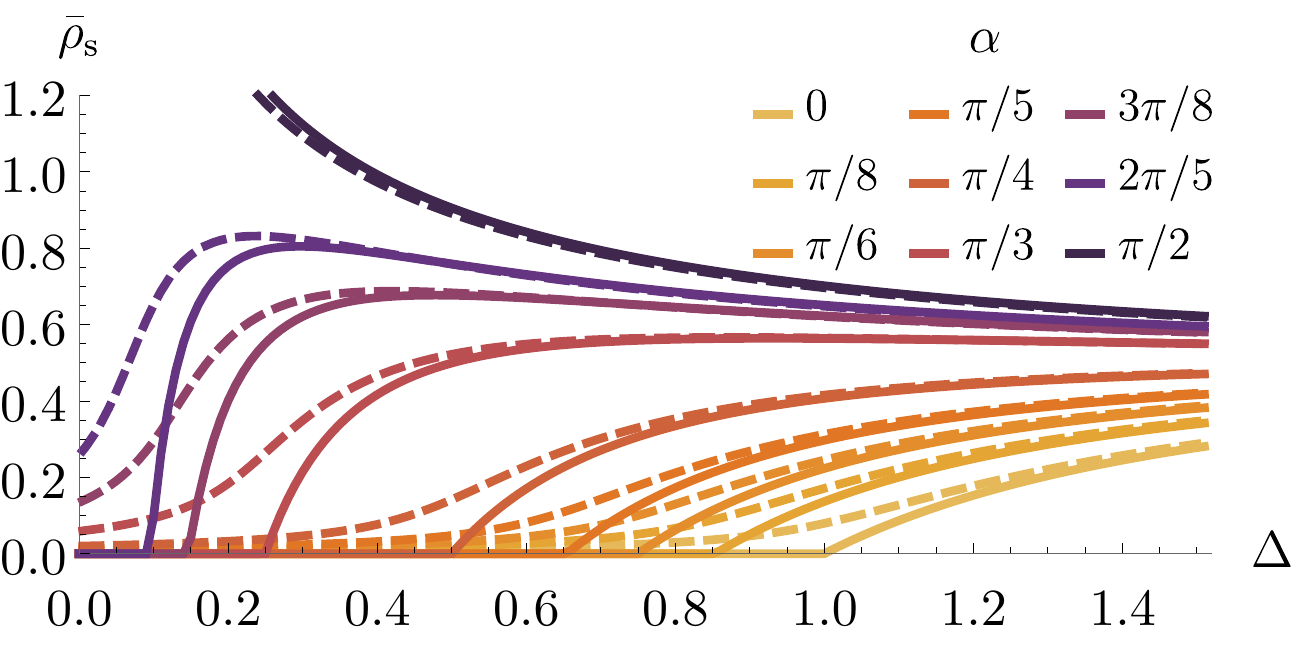}
\caption{Surface density of Dirac surface states ($b=0$, solid curves) and  Fermi arcs ($b=\Lambda/10$, dashed curves), as a function of disorder strength $\Delta$, for several mixing angles between the chiral degrees of freedom $\alpha\in[0,\pi/2]$. The Dirac surface states become metallic above the critical value $\Delta_{\rm s}^* = (\cos \alpha)^2$. Weyl semimetals avoid this critical point.}
\label{fig:FADoS}
\end{figure}

\begin{figure}[b!]
\centering
\includegraphics[scale=1]{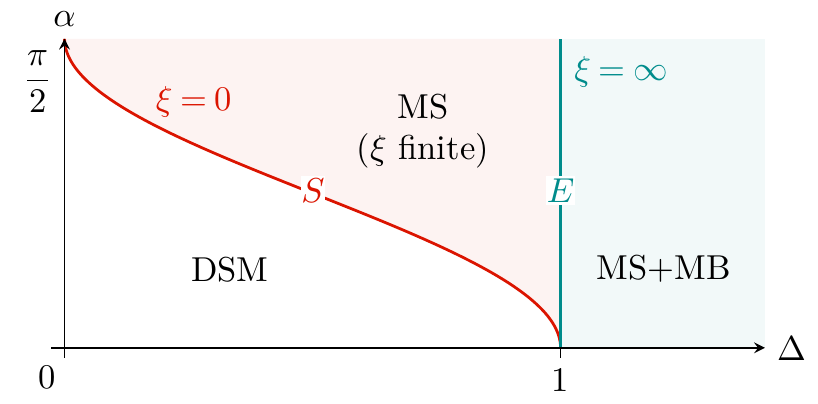}
\caption{Phase diagram in the $(\Delta,\alpha)$ plane of 
Dirac semimetals hosting single-cone surface states (DSM). Beyond the surface critical line $S$ : $\Delta_{\rm s}^* = (\cos \alpha)^2$, metallic eigenstates populate the surface (MS). Beyond the extraordinary line $E$ : $\Delta_{\rm b}^*=1$, the bulk 
becomes metallic as well (MS+MB).}
\label{fig:phase_diagram}
\end{figure}
The LSCBA leads to the surface density of Dirac surface states shown as a function of disorder strength~$\Delta$ and for various {surface Fermi velocities $v_0=\cos \alpha$ in Fig.~\ref{fig:FADoS}. Except for nondispersing flat bands ($\alpha=\pi/2$), eigenstates do not populate the Fermi node until a critical disorder strength $\Delta_{\rm s}^* = v_0^2$, in agreement with dimensional analysis.
The corresponding phase diagram, shown in Fig.~\ref{fig:phase_diagram}, resembles the one for semi-infinite spin systems~\cite{lubensky_critical_1975, lubensky_critical_1975_a}, except that there is no ordinary transition, where the surface and the bulk develop an order parameter simultaneously. At the surface critical line $\Delta_{\rm s}^*(\alpha)$, the boundary turns into a metallic state, while the bulk remains a semimetal. The bulk undergoes a transition only at $\Delta_{\rm b}^*>\Delta_{\rm s}^*$, when the surface is already metallic; this is known as an extraordinary transition. In addition, the surface and extraordinary critical lines merge at the special point $(\alpha=0,\Delta=1)$.  
In contrast to spin systems, the surface transition can only exist on the boundary of three-dimensional semimetals, but not in infinite two-dimensional Dirac materials.

\begin{figure}[t!]
\centering
\includegraphics[scale=0.722]{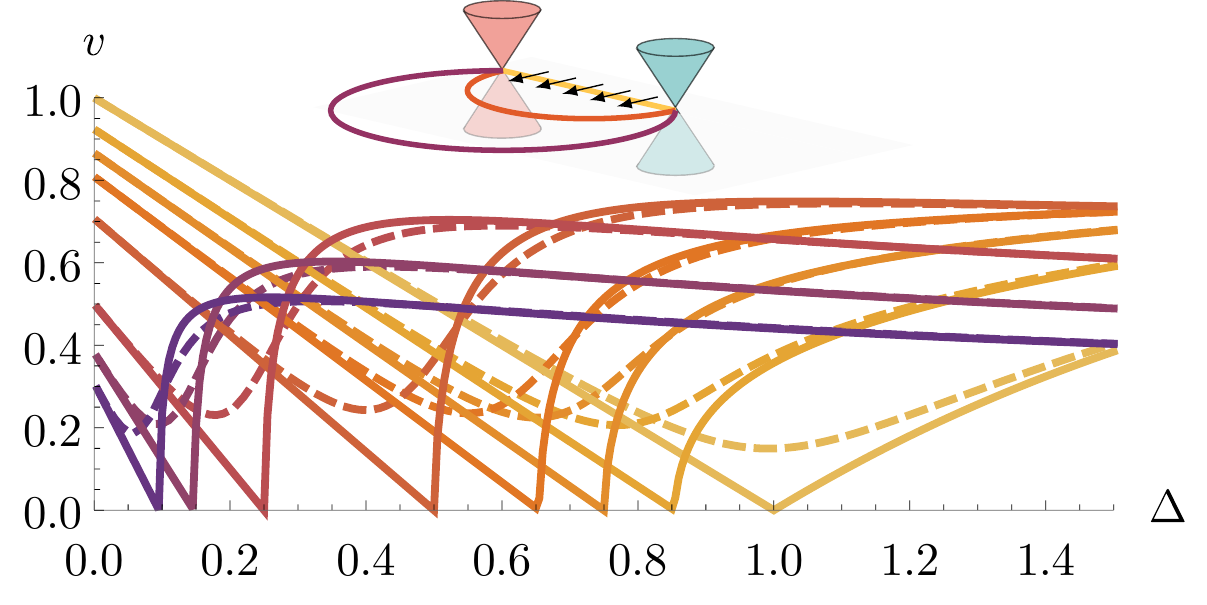}
\caption{Group velocity $v$ as a function of disorder strength $\Delta$, for several $\alpha\in [0,\pi/2]$ (see legend of Fig.~\ref{fig:FADoS}). The solid curves are for Dirac surface states ($b=0$); the dashed curves for Fermi arcs ($b=\Lambda/10$). Inset: the group velocity is locally orthogonal to the arcs. }
\label{fig:vg}
\end{figure}

The group velocity also reveals a critical behavior at the surface transition. Using its definition $\bm{v} = \partial_{\bm{k}} \epsilon_{\rm s}$~\cite{Wilson:2018}, where $\epsilon_{\rm s}(\bm{k})$ denotes the surface relation dispersion, we express it as \cite{Supplementary}  
\begin{equation}\label{eq:v} 
\bm{v} = \bm{v_0}[1-\partial_\epsilon(\text{Re} \Sigma)(\epsilon_{\rm F}(z=0),z=0)]^{-1}
\end{equation}
at the surface nodal level, where $\bm{v_0}$ is the group velocity of the clean Dirac surface states. We compute $\partial_\epsilon(\text{Re}\Sigma)$ 
by solving Eq.~\eqref{eq:SCBA} numerically for energies $\epsilon$ around $\epsilon_{\rm F}(0)$. Figure~\ref{fig:vg} shows that at surface criticality, where the dispersion relation reads $\epsilon_{\rm s} \propto k^{z_{\rm s}}$, the group velocity vanishes like $v \propto |\Delta-\Delta^*_{\rm s}|^{\nu(z_{\rm s}-1)}$ with the surface dynamical exponent $z_{\rm s}=2$.

\emph{\label{sec:DFA} Effect of disorder on Fermi arcs.} --
As shown in Fig.~\ref{fig:FADoS}  a nonzero separation $b$ between the surface-projected nodes generates a nonzero surface density $\bar{\rho}_{\rm s} \propto b^2$ for arbitrary weak disorder. This smooths out the sharp surface transition.
The penetration length of a surface state has a minimum at the middle of the Fermi arc and is infinite at the nodal points~\cite{gorbar_origin_2016}.  With increasing disorder strength the Fermi arcs broaden. The minimal penetration length grows with disorder and diverges at the extraordinary transition, which reflects dissolving of the surface states into the bulk~\cite{gorbar_origin_2016}.
The group velocity computed  for the Fermi arc states at the junctions with the nodal points is shown in Fig.~\ref{fig:vg}}. It always remains finite, which also indicates that the surface transition is avoided.

\emph{\label{sec:summary} Conclusion and Outlook.} -- 
We have studied the full phase diagram of disordered semi-infinite relativistic semimetals as a function of boundary conditions and disorder strength. To that end we calculated the spatially resolved density of states and the surface group velocity using a local self-consistent Born approximation.   
We have shown that with increasing the strength of disorder,
Dirac semimetals hosting single-cone surface states undergo a surface phase transition from a semimetal to a surface diffusive metal, followed by an extraordinary transition to a bulk diffusive metal. For Weyl semimetals this surface transition is smoothed out  
due to the finite extension of the Fermi arcs or Fermi rays. We found that within the LSCBA the critical exponents at the surface and special transitions  are identical to those computed using the SCBA for the bulk transition. However, we expect that their actual values are different similar to the Anderson transition in a semi-infinite systems \cite{Subramaniam2006}. The multifractality of critical surface states is also expected to differ from that in the bulk~\cite{Brillaux:2020,Syzranov:2016a}.   

\emph{\label{sec:acknow} Acknowledgments.} -- 
We would like to thank David Carpentier, Lucile Savary and Ilya Gruzberg for inspiring discussions.
We acknowledge support from the French Agence Nationale de la Recherche  by the Grant No.~ANR-17-CE30-0023 (DIRAC3D), ToRe IdexLyon breakthrough program, and ERC under the European Union's Horizon 
2020 research and innovation program (project TRANSPORT N.853116).

\newpage

\bibliographystyle{apsrev4-1}
%

\widetext
\setcounter{equation}{0}
\renewcommand{\theequation}{\thesection\arabic{equation}}
\renewcommand{\thesection}{\Alph{section}}
\renewcommand{\thesubsection}{\arabic{subsection}} 
\begin{center}
\MakeUppercase{\textbf{\Large Supplemental Material}}
\end{center}

\section{Green's function of the clean semi-infinite semimetal}
\label{sec:Green_function}

The  Green's function $G_0$ of a clean semi-infinite system can be Fourier-transformed in the two directions $(x,y)$ parallel to the surface, but not in the transverse direction $z$ due to the absence of translational invariance. We can thus express $G_0$ in a mixed momentum-real space, as a function of the parallel components $k_x$ and $k_y$ of the wavevector, and of two distances $(z,z')$ to the boundary. To simplify the notation we will omit the explicit dependence of $G_0$ on $k_x$ and $k_y$. The Green's function can be found from the equation  
\begin{equation}\label{eqS:def_G0} 
[H_0 - \epsilon \tau_0 \sigma_0 ]G_0(\epsilon,z,z') = \delta(z-z') \tau_0 \sigma_0,
\end{equation}
where the bulk Hamiltonian $H_0 = \tau_z\sigma_x k_x + \tau_z\sigma_y k_y + i \tau_z\sigma_z \partial_z  + \tau_0 \sigma_x b + \tau_0 \sigma_z b_z$ is expressed in mixed momentum-real space. The Green's function has to satisfy the boundary condition $MG_0(\epsilon,0,z')=G_0(\epsilon,0,z')$.
A general solution to Eq.~\eqref{eqS:def_G0} can be written as a sum of two terms: a particular solution 
for the infinite system, and a solution of the corresponding  homogeneous equation. 
The part of the solution which is off-diagonal in the pseudospin sector reads~\cite{faraei_greens_2018}
\begin{equation}\label{eqS:G0_off} 
{(G_0)}^{\bar{\sigma} \sigma}_{\chi \chi'}(\epsilon,z,z') = \dfrac{\chi(k_x^\chi + i\sigma k_y^\chi)}{2(q_\chi+i\chi b_z)} \left[ \delta_{\chi \chi'} e^{-(q_\chi +i \chi b_z)|z-z'|} - {(A_0)}^{\bar{\sigma} \sigma}_{\chi \chi'} e^{-(q_\chi + i\chi b_z)(z+z')]}  \right],
\end{equation}
where $\sigma$ and $\bar{\sigma} = -\sigma$ are pseudospin components, $\chi,\chi'$ are two independent chiral components, $\bm{k}^{\chi} = \bm{k} + \chi b \, \bm{e_x}$ is the momentum measured from the surface-projected node of chirality $\chi$, and $q_\chi^2 = (\bm{k^\chi})^2 - \epsilon^2$. 
Like the Hamiltonian, the Green's function is a four-times-four matrix, defined in the product space of  the pseudospin and chiral sectors. 
The components 
which are diagonal in pseudospin can be found thanks to the relation
\begin{equation}
{(G_0)}^{\sigma \sigma}_{\chi \chi'}(\epsilon,z,z') = \dfrac{\epsilon- i\chi\sigma \partial_z + \sigma b_z}{\chi(k_x^\chi + i\sigma k_y^\chi)} {(G_0)}^{\bar{\sigma} \sigma}_{\chi \chi'}(\epsilon,z,z').
\end{equation}
The coefficient $A_0$ of Eq.~\eqref{eqS:G0_off} is fixed by the boundary conditions.
 For the $M_1$ boundary condition, only the diagonal chiral components are nonzero:
\begin{equation}\label{eqS:AM1-0} 
\left[ \left(A_0\right)_{\chi \chi'}^{\bar{\sigma} \sigma} \right]_1 =  \dfrac{\epsilon-i\chi\sigma q_{\chi} + 2\sigma b_z -\chi e^{-i\sigma\theta_\chi}(k_x^\chi + i\sigma k_y^\chi)}{\epsilon+i\chi\sigma q_{\chi }-\chi e^{-i\sigma\theta_\chi}(k_x^\chi + i\sigma k_y^\chi)} \delta_{\chi \chi'},
\end{equation}
where $\theta_{\pm}$ are the two angles that parametrize $M_1$. For the $M_2$ boundary condition, the chiral components are coupled, and $A_0$  takes a more complex form,
\begin{equation}\label{eqS:AM2-0} 
\left[ \left(A_0\right)_{\chi \chi'}^{\bar{\sigma} \sigma} \right]_2 =   1- \dfrac{ 2i\chi \sigma e^{i\chi \theta_{\bar{\sigma}}} q_{\chi}(k_x^{\bar{\chi}} + i\sigma k_y^{\bar{\chi}}) }{D_{\chi \chi'}^{\bar{\sigma} \sigma} },
\end{equation}
where the denominator reads
\begin{equation}\label{eqS:Dpole0} 
D_{\chi \chi'}^{\bar{\sigma} \sigma} = e^{i\chi \theta_{\bar{\sigma}}} (\epsilon  + i\chi \sigma q_{\chi})(k_x^{\bar{\chi}} + i\sigma k_y^{\bar{\chi}}) + e^{i\chi \theta_{\sigma}} (\epsilon  + i\bar{\chi} \sigma q_{\bar{\chi}} )(k_x^{\chi} + i\sigma k_y^{\chi}).
\end{equation}

\setcounter{equation}{0}
\section{Surface states of the clean system}
\label{sec:surface_states}

The Green's function for the clean system of Eq.~\eqref{eqS:G0_off} admits a pole proportionnal to $q_\chi + i \chi b_z$. This pole determines the bulk dispersion relations $\epsilon^{\chi}_{\rm b}(\bm{k}) = \sigma \sqrt{(k^\chi)^2 + b_z^2}$ for the two nodes of chirality $\chi=\pm 1$, where the pseudospin $\sigma=\pm 1$ dictates the particle or hole side of the energy band. However, the coefficient $A_0$ of the excess contribution to the Green's function, which appears only in the presence of a boundary, admits a different pole. This other pole determines the dispersion relation of the surface-localized eigenstates, and depends on the boundary conditions. We determine the properties of these surface states for all possible boundary conditions: $M_1$, $M_2$ and all combinations thereof which respect the unitarity and the vanishing of the transverse current.

\subsection{\texorpdfstring{$M_1$}{M1} boundary condition}
\label{secS:M1BC}

 The pole $D_1$ for the $M_1$ boundary condition is the denominator of Eq.~\eqref{eqS:AM1-0} :
\begin{equation}\label{eqS:D1} 
D_1(\epsilon,\bm{k}) = \epsilon + i\chi q_{\chi} -\chi e^{-i\sigma\theta_\chi}\left(k_x^\chi + i k_y^\chi\right).
\end{equation}
There are two independent relation dispersions $\epsilon_{\rm s}^\chi(\bm{k})$ : one for each node of chirality $\chi=\pm 1$. Thus the properties of the surface states arising from the $M_1$ boundary condition are independent of the relative position of the surface-projected nodes, and in particular of whether the semimetal is binodal Weyl or single-node Dirac. Solving the equation $D_1(\epsilon_{\rm s}^\chi,\bm{k})=0$, we find~\cite{faraei_greens_2018}
\begin{subequations}
\begin{align}[left = \empheqlbrace\,]
\epsilon_{\rm s}^\chi & = \chi k_\chi \cos(\phi_\chi-\theta_\chi), \label{eqS:epsilon1}\\
q_\chi & = k_\chi \sin(\phi_\chi-\theta_\chi),
\end{align}
\end{subequations}
where $k_\chi$ and $\phi_\chi$ are respectively the norm and the angle of the wavevector $\bm{k^\chi}$ measured from the node of chirality~$\chi$. The pseudospin component $\sigma$ does not intervene : the surface energy band is two-fold degenerate. The corresponding eigenstates $\psi_1$ can be found by solving the eigenvalue equation $H_0 \psi_1(x,y,z) = \epsilon \psi_1(x,y,z)$ with $M_1 \psi_1(x,y,0)= \psi_1(x,y,0)$. They can be written as superpositions of the two orthogonal eigenstates for each node, 
\begin{equation}
\psi_1(x,y,z) = N_1 e^{-i(xk_x + yk_y)}\left[C_1 e^{-zq_+} \left(\begin{array}{c}1\\e^{i\theta_+}\\0\\0\end{array}\right)  + C_2 e^{-zq_-} \left(\begin{array}{c}0\\0\\1\\e^{i\theta_-}\end{array}\right)\right],
\end{equation}
where the normalization constant $N_1$ is found by imposing $\int |\psi|^2 = 1$ over the $z>0$ half-space, and $C_1$, $C_2$ are two constants. Such eigenstates are normalizable, and describe surface-localized plane waves provided that $q_\chi>0$, which selects for each node the half-plane $\phi_\chi-\theta_\chi \in [0,\pi]$. 

At the Fermi level, Eq.~\eqref{eqS:epsilon1} vanishes and the surface band structure reduces to the half-lines $\phi_\chi = \theta_\chi+\pi/2$. Such band structures are called Fermi rays. They are orthogonal to the unitary vectors $\bm{e_\chi}=(\cos \theta_\chi,\sin \theta_\chi,0)$. The eigenstates are more strongly bound to the surface as we move on the rays away from the nodes. Indeed, their amplitude decays exponentially into the volume of the system over a distance $1/q_\chi=1/k_\chi$.

\subsection{\texorpdfstring{$M_2$}{M2} boundary condition}
\label{secS:M2BC}

 The pole $D_2$ for the $M_2$ boundary condition is the denominator of Eq.~\eqref{eqS:AM2-0} :
\begin{equation}\label{eqS:D2} 
D_2(\epsilon,\bm{k}) = -e^{-i\chi\alpha} \left(\epsilon + i \chi\sigma q_\chi\right)\left(k_x^{\bar{\chi}} + i \sigma k_y^{\bar{\chi}}\right) + e^{i\chi\alpha} \left(\epsilon - i \chi \sigma q_{\bar{\chi}}\right)\left(k_x^{\chi} + i \sigma k_y^{\chi}\right),
\end{equation}
where we assumed that $\theta_+ = \alpha$ and $\theta_- = \pi-\alpha$, so that the surface projection of the nodes are aligned in the $x$ direction. In contrast to Eq.~\eqref{eqS:D1}, the pole for the $M_2$ boundary condition couples the parameters associated to both Weyl nodes, e.g. the momenta $k_+$ and $k_-$, as a direct consequence of chirality mixing. Hence, we expect that a continuous band structure emerges from one node and terminates on the other. Solving $D_2(\epsilon_{\rm s},\bm{k})=0$, we find indeed a single dispersion relation $\epsilon_{\rm s}(\bm{k})$ for both chiralities,
\begin{subequations}
\begin{align}[left = \empheqlbrace\,]
\epsilon_{\rm s} & = k_\chi \cos \beta_\chi,\label{eqS:epsilon2}\\
q_\chi & = k_\chi \sin \beta_\chi,
\end{align}
\end{subequations}
where the angles $\beta_\chi$ are completely determined by the relation $\beta_+ + \beta_- = 2\alpha + \phi_+ - \phi_-$, along with the condition $k_{+} \cos \beta_+ =k_{-} \cos \beta_- $ of Eq.~\eqref{eqS:epsilon2}. Like for Fermi rays, the pseudospin leads to a two-fold degeneracy of the energy levels. In addition, the band structure depends on the separation between the surface projections of the nodes. We must distinguish qualitatively the cases for which this separation vanishes, from those for which it is nonzero.\\

\begin{figure}[b!]
\newcommand{\HeiGht}{-0.8cm}\subfigure{\includegraphics[scale=0.361]{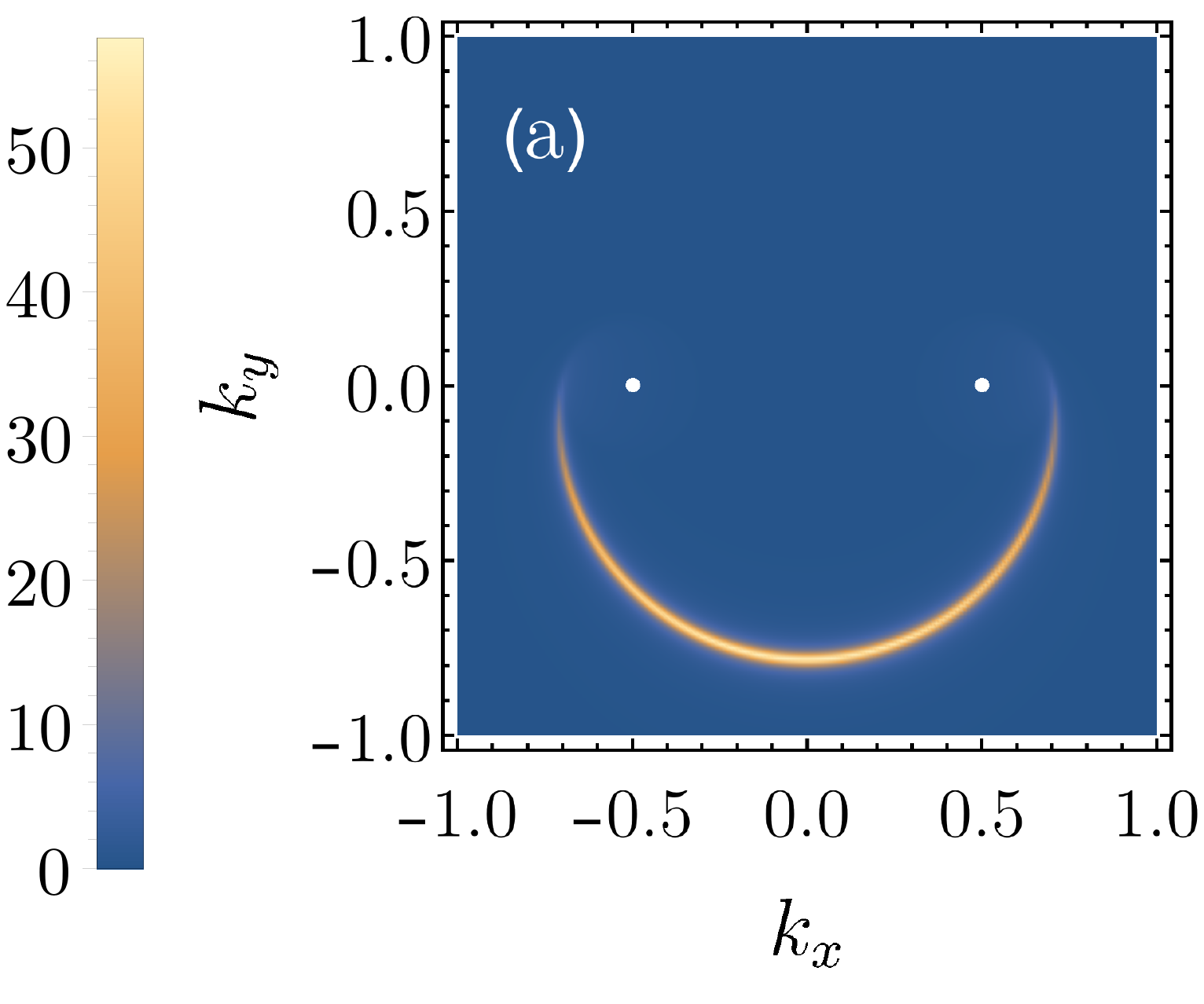}}\hspace{\HeiGht}~~~\subfigure{\includegraphics[scale=0.361]{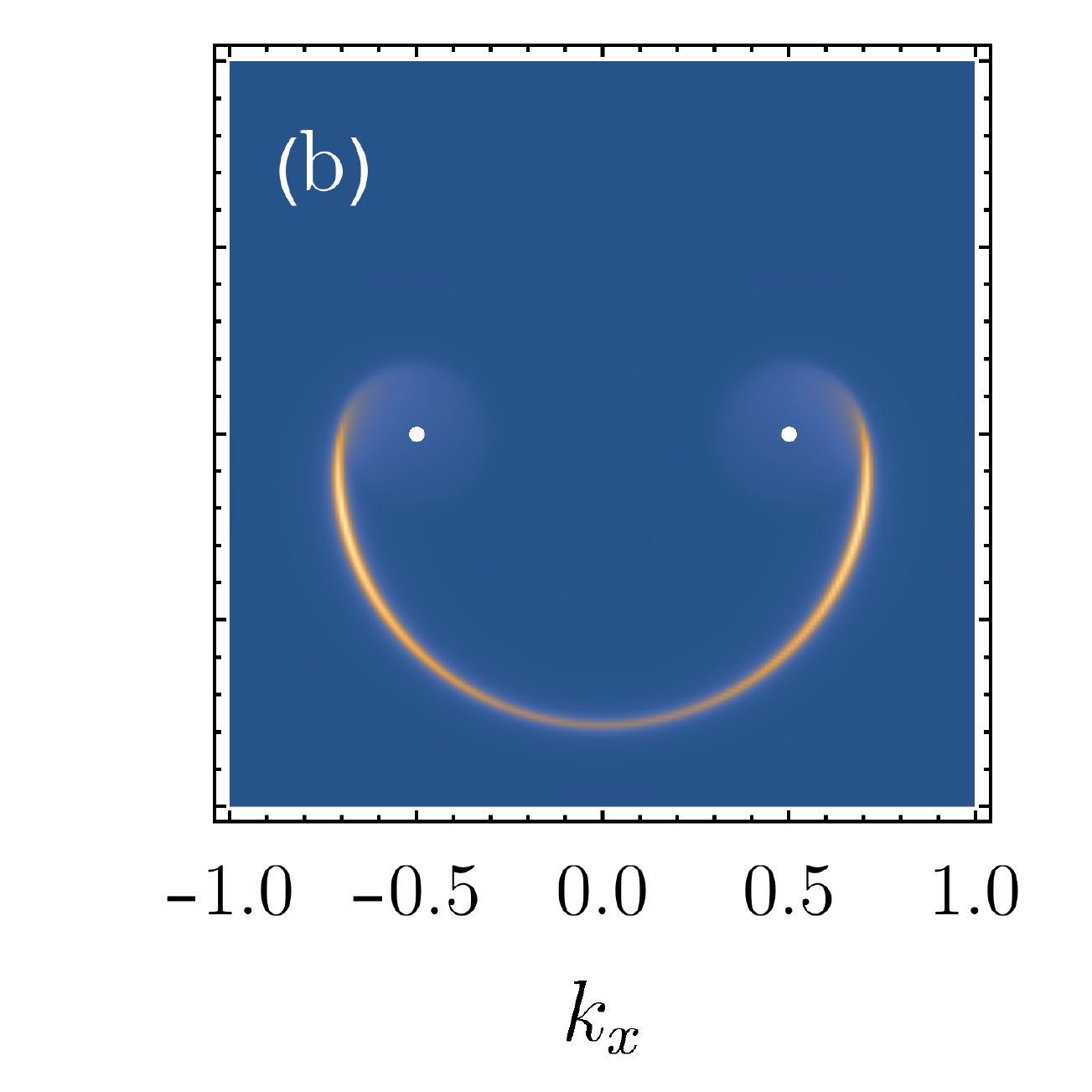}}\hspace{\HeiGht}\subfigure{\includegraphics[scale=0.361]{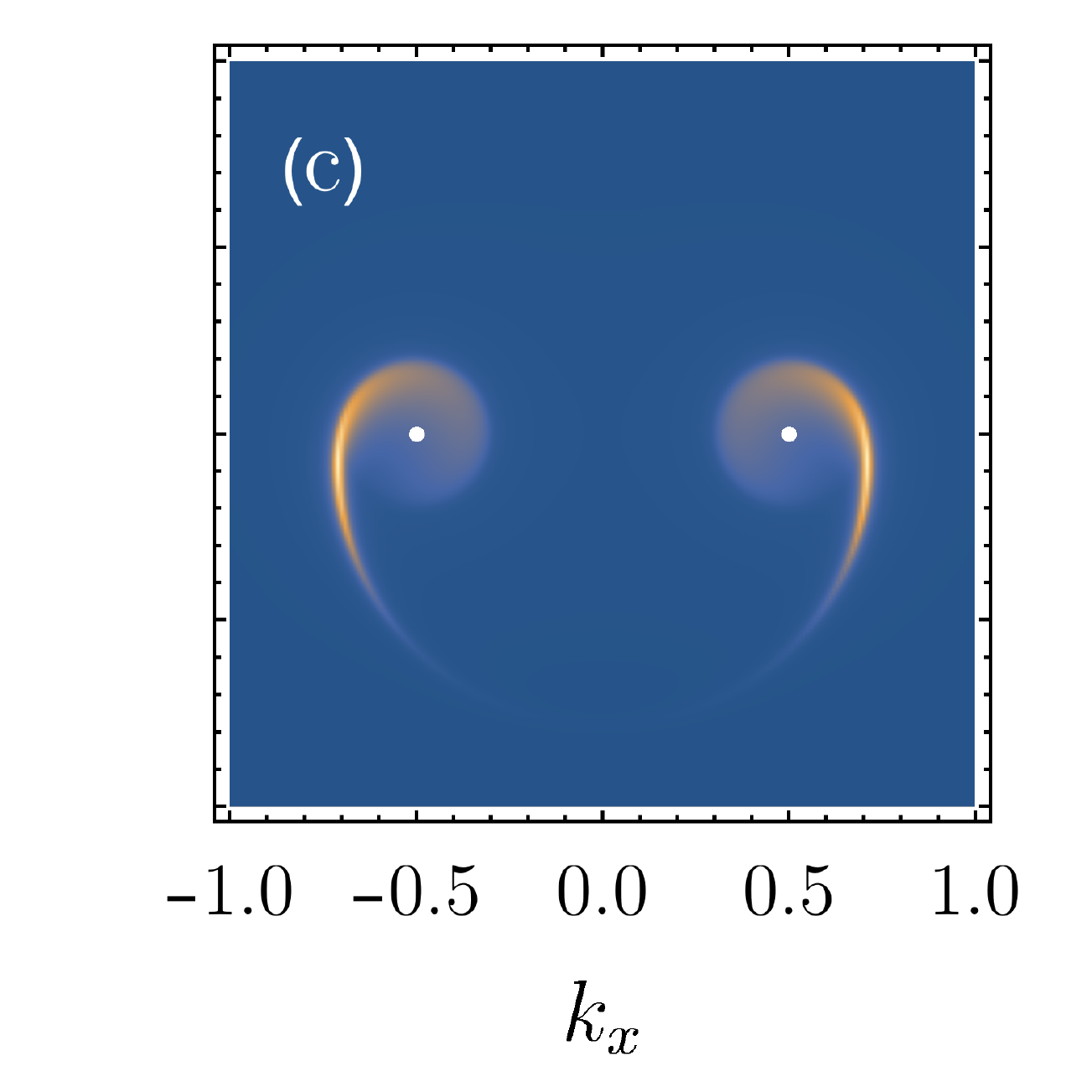}}\hspace{\HeiGht}\subfigure{\includegraphics[scale=0.361]{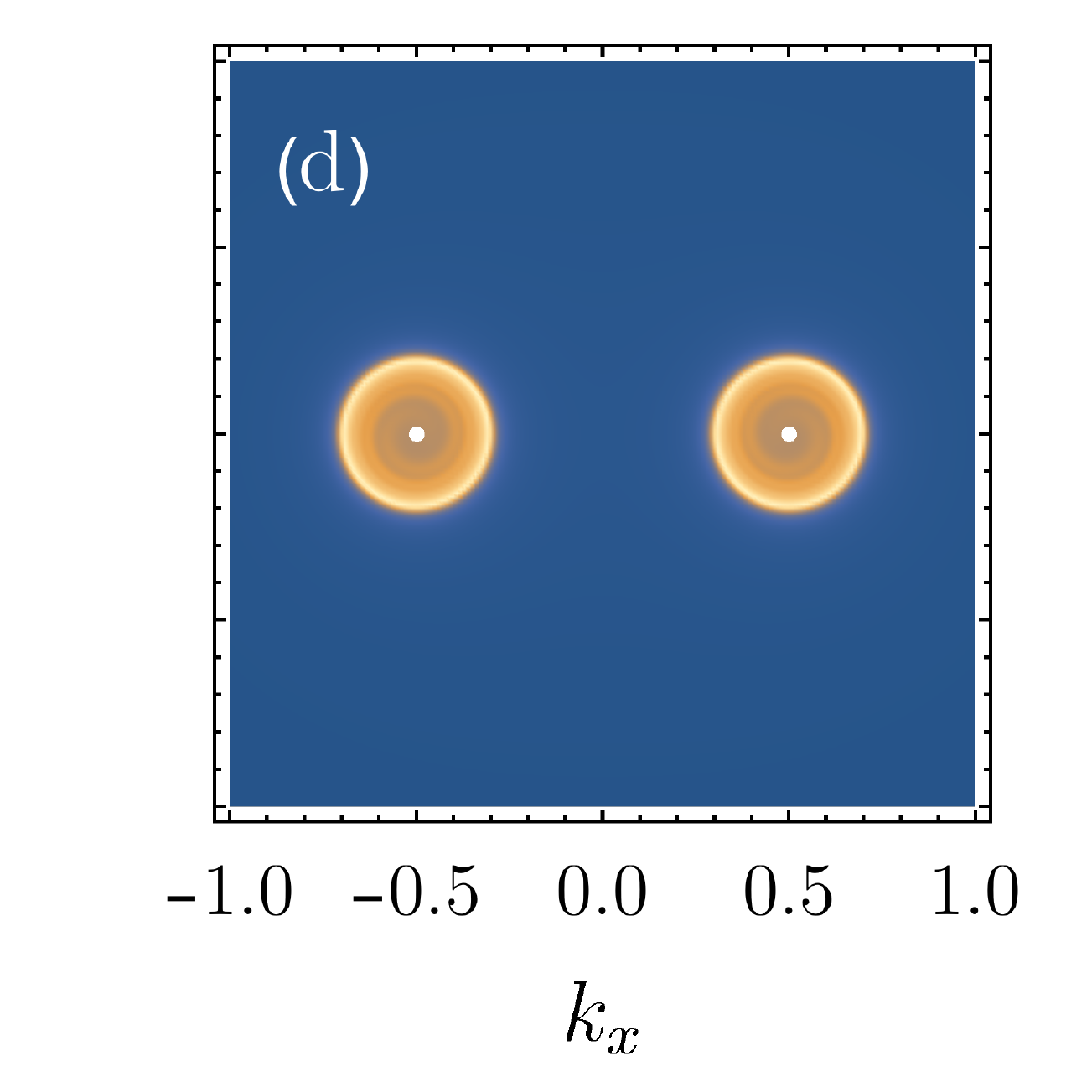}}
\caption{Spectral density map $A(k_x,k_y,\epsilon,z)$ of a Fermi arc with curvature $\alpha=\pi/4$ over the Brillouin zones projected to a slab of fixed distance $z$ from the surface, at energy $\epsilon=0.2$, where all energies and momenta are normalized by the cut-off $\Lambda$. (a) $\Lambda z =0$, (b) $\Lambda z =2$, (c) $\Lambda z =5$, and (d) $\Lambda z =\infty$.  The nodes (white dost) are separated by a momentum $2b=1$ along the $k_x$ axis. The Fermi arcs penetrate over some finite depth in the bulk. The scattering rate is set to $\Gamma=0.01$ and broadens the band structures. The wavefunction decays like $e^{-qz}$ in the bulk, where $q$ is largest at the middle of the arc, and vanishes at the contact with the cones.}
\label{fig:spectral_density} 
\end{figure}

\paragraph{Weyl semimetal without superposition of surface-projected nodes} For a nonzero half-separation $\bm{b}=b \bm{e_x} + b_z \bm{e_z}$ between the nodes, the Fermi level at the boundary ($\epsilon_{\rm s}=0$) hosts a Fermi arc, whose parametric equation reads
\begin{equation}\label{eqS:Arc} 
\left(\dfrac{k_x^- + ik_y^-}{k_-}\right) = -e^{2i\alpha} \left(\dfrac{k_x^+ + ik_y^+}{k_+} \right).
\end{equation}
Equation~\eqref{eqS:Arc} describes a circle of diameter $d_{\rm FA} = b/\sin \alpha$ and center $\bm{k_{\rm FA}} = -2b\cot(2\alpha) \bm{e_y}$. But only normalizable eigenstates are physical : the requirement $q_\chi>0$, along with $\epsilon_{\rm s}=0$, imposes $\beta_+=\beta_-=\pi/2$, which restricts the arc to the half-plane for which $\phi_+-\phi_- = \pi-2\alpha \in [0,\pi]$. The other half-plane can be explored for $\alpha \in[\pi/2,\pi]$, or by a mirror symmetry $y\mapsto -y$. The perimeter of this Fermi arc $L_{\rm FA} = 4b\alpha / \sin(2\alpha)$ also gives its density of states. Notice that the density of the Fermi arc differs from the surface density computed in the main text, since the arcs leak into the bulk over a certain distance, which diverges close to the nodes. However, the two should manifest the same behavior under disorder. A compelling way to picture the band structure is through the
spectral density
\begin{equation}
A(k_x,k_y,\epsilon,z) = \dfrac{\text{Im}[\text{Tr}G(k_x,k_y,\epsilon,z)]}{\pi},
\end{equation}
where Im denotes the imaginary part.
The spectral density is a function of the position ($k_x$,$k_y$) on the two-dimensional projection of the Brillouin zone over a slab of fixed distance $z$ to the surface and fixed energy $\epsilon$. The band structure appears as a singularity of the spectral map. Examples of spectral density maps are proposed in Fig.~\ref{fig:spectral_density} above the Fermi level, where the arcs dissolve in the bulk Weyl cones in a spiraling fashion.

\paragraph{Single Dirac cone or Weyl semimetal with superposition of surface-projected nodes} For $\bm{b}=\bm{0}$, the surface energy band of Eq.~\eqref{eqS:e2} and \eqref{eqS:q2} reduce to 
\begin{subequations}
\begin{align}[left = \empheqlbrace\,]
\epsilon_{\rm s} & = k \cos \alpha,\label{eqS:e2}\\
q & = k \sin \alpha.\label{eqS:q2}
\end{align}
\end{subequations}
The surface-localized eigenstates $\psi_2$ thus distribute along the semi-cone $\epsilon_{\rm s}>0$ with Fermi velocity $v_0 = \cos\alpha$. Their expression reads
\begin{equation}
\psi_2(x,y,z) = N_2 e^{-i(xk_x + yk_y)} e^{-zk\sin \alpha} \left(\begin{array}{c}e^{-i(\alpha+\phi)}\\1\\e^{-i\phi}\\-e^{-i\alpha}\end{array}\right),
\end{equation}
up to a normalization constant $N_2$. At the Fermi level, the band structure reduces to a Fermi node. Following Ref.~\cite{ShtankoLevitov2018}, we call these surface eigenstates \emph{Dirac surface states}, though they are not specific to Dirac semimetals but also emerge whenever two Weyl nodes are projected onto the same point in the surface Brillouin zone.

\subsection{Mixing between boundary conditions}
\begin{figure}[b!]
\newcommand{\HeiGht}{-0.8cm}\subfigure{\includegraphics[scale=0.361]{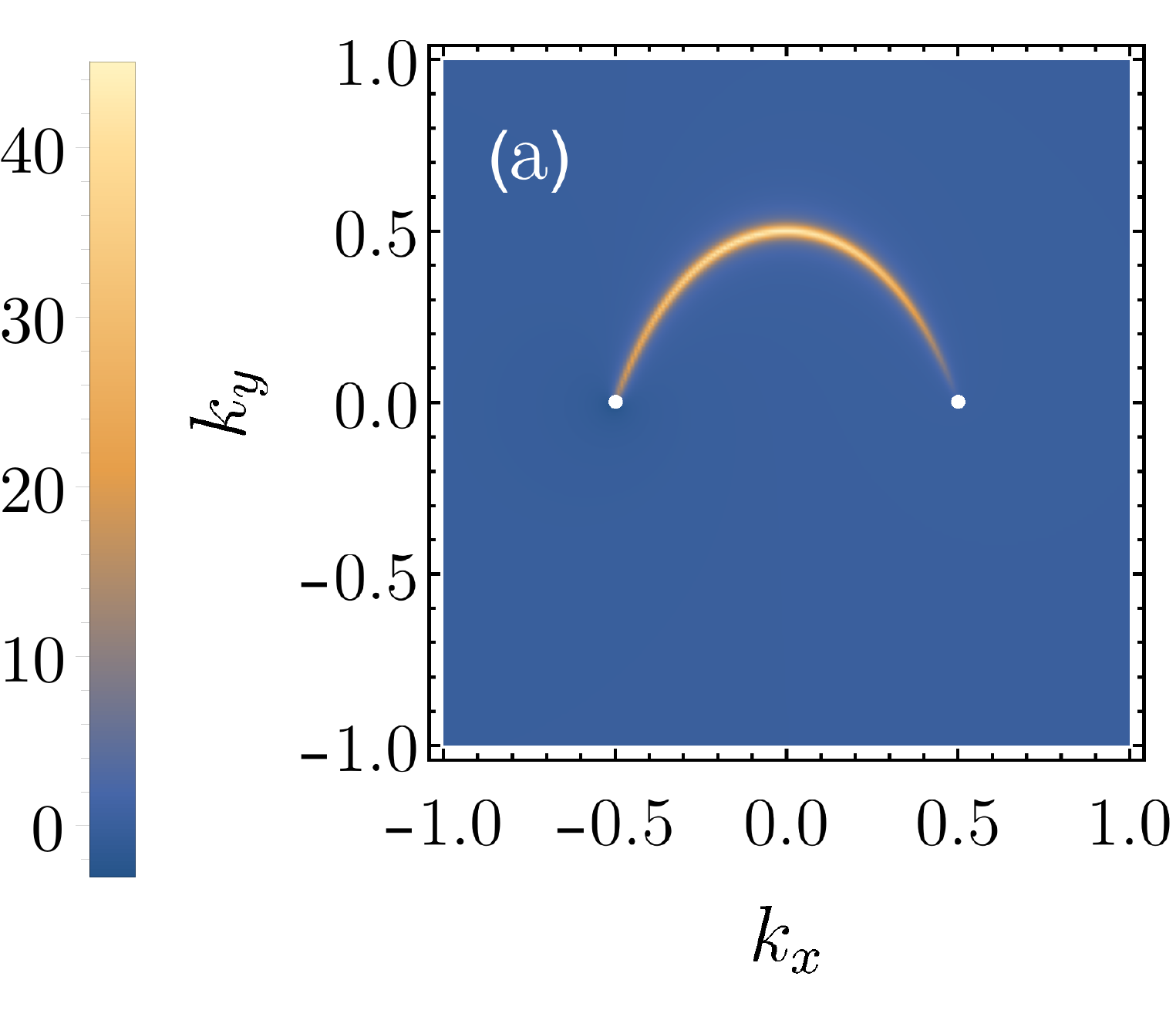}}\hspace{\HeiGht}~~~\subfigure{\includegraphics[scale=0.361]{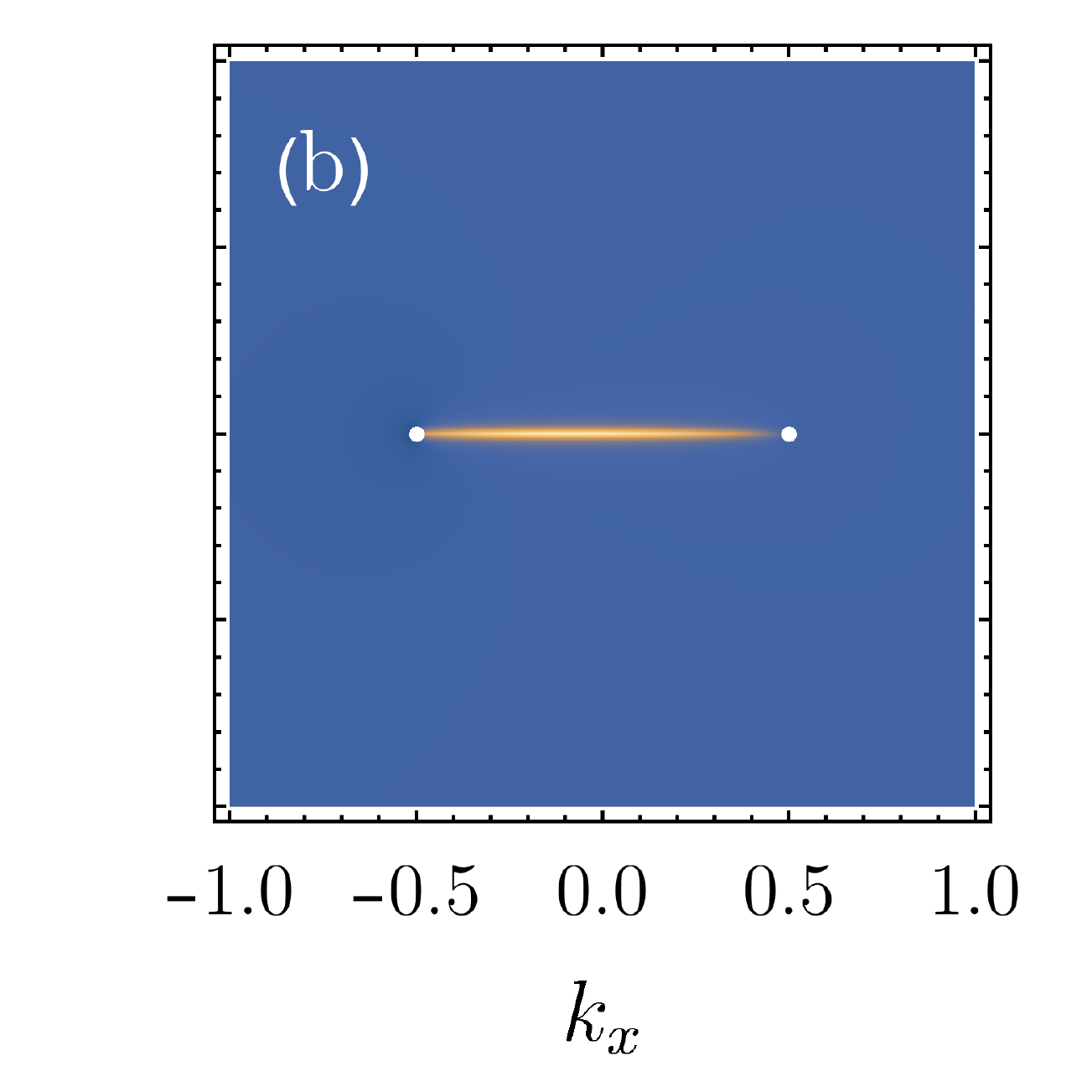}}\hspace{\HeiGht}\subfigure{\includegraphics[scale=0.361]{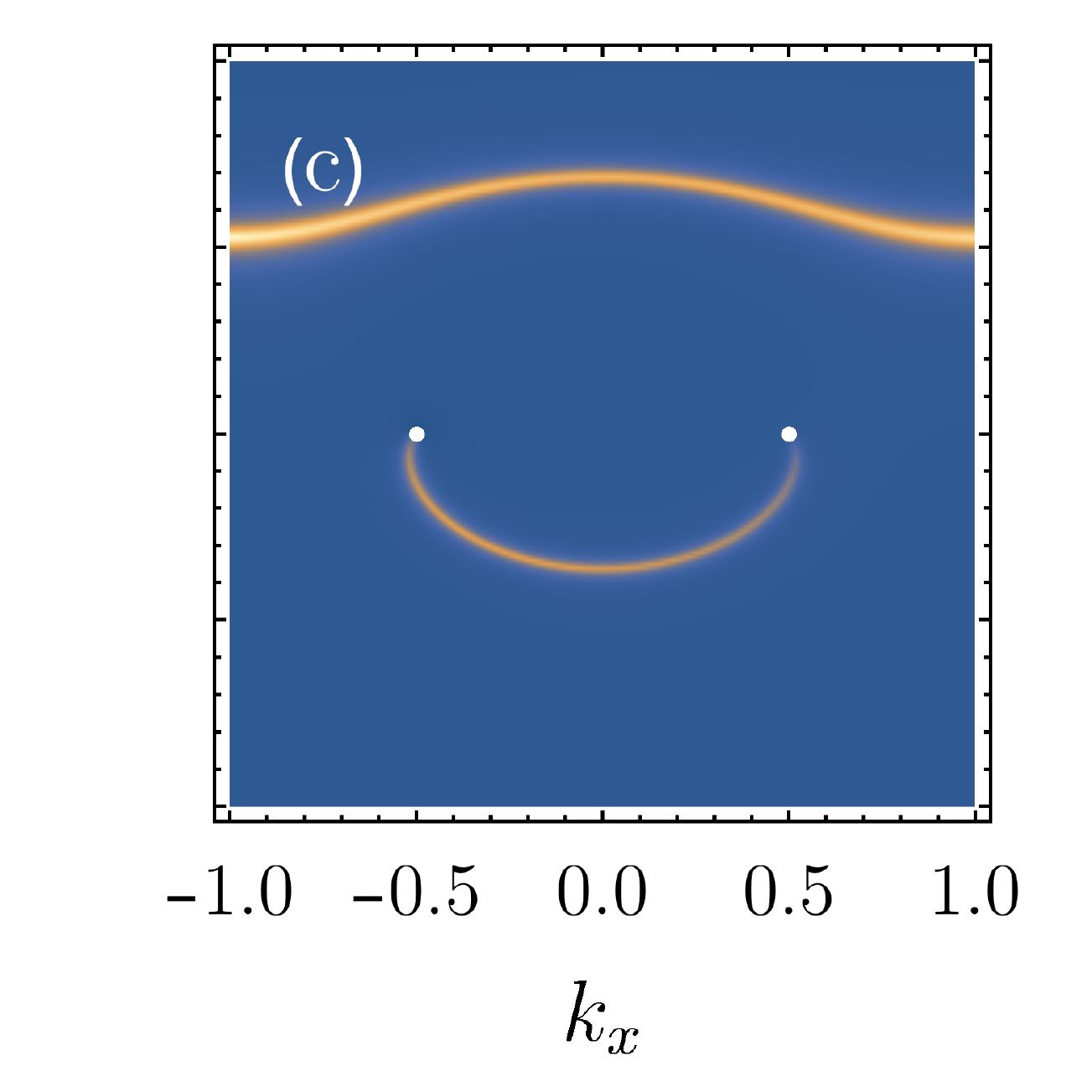}}\hspace{\HeiGht}\subfigure{\includegraphics[scale=0.361]{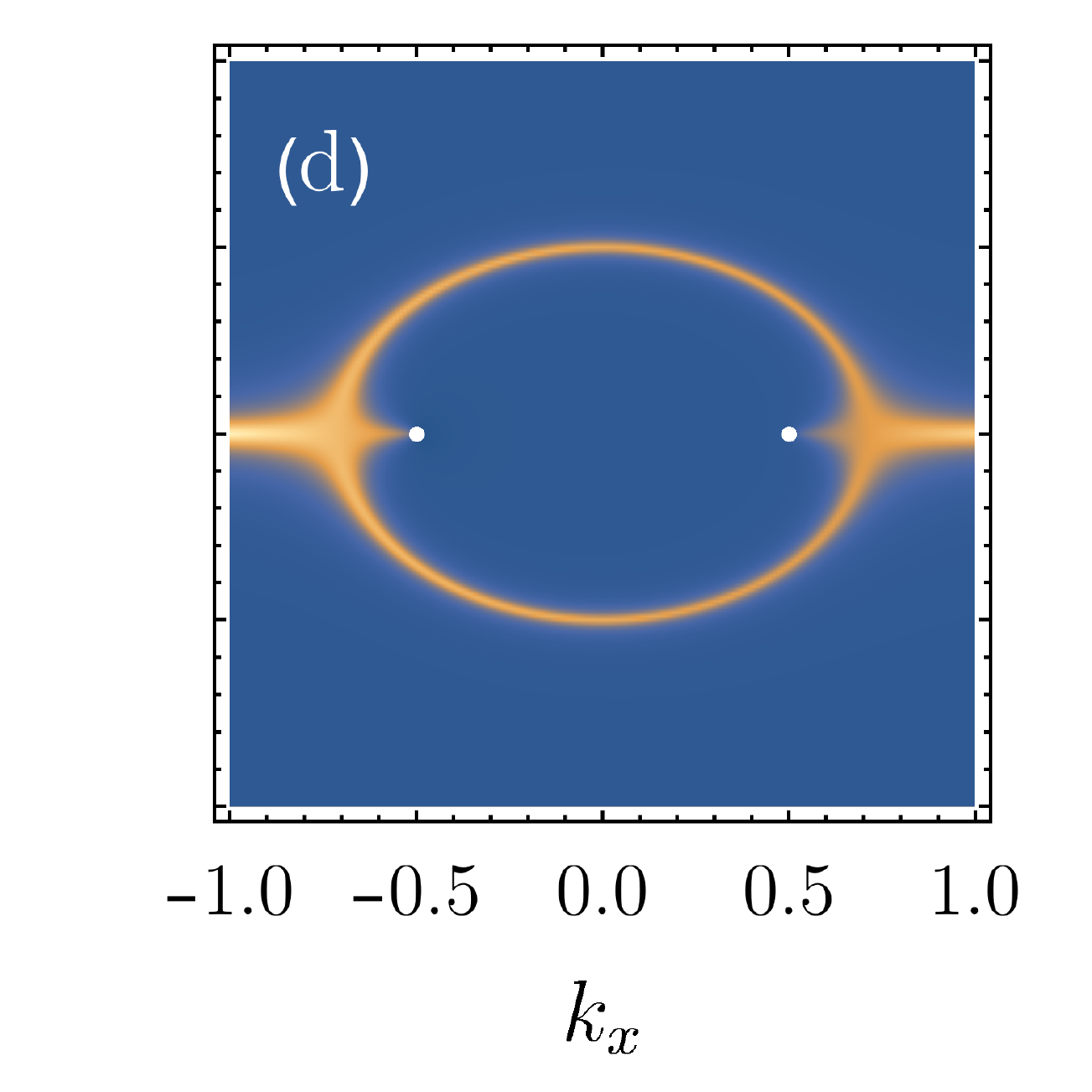}}
\caption{Surface spectral density map at the Fermi level $A(k_x,k_y,0,0)$, for a mixing of the boundary conditions $M_1$ and $M_2$ with equal weight ($\varphi=\pi/4$) preserving the mirror symmetry $x\mapsto -x$ ($\theta_+ + \theta_-=0$). The parameter $\alpha$ is set to (a) $\alpha=0$, (b) $\alpha = \pi/4$, (c) $\alpha=9\pi/20$, and (d) $\alpha=\pi/2$. The scattering rate is set to $\Gamma = 0.05 \Lambda$. 
}
\label{figS:spectral_density} 
\end{figure}

We distinguished two types of boundary conditions : $M_1(\theta_\pm) = \tau_+ (\bm{\sigma} \cdot \bm{e_{+}}) + \tau_{-}(\bm{\sigma} \cdot \bm{e_{-}})$, which is parametrized by two independent angles $\theta_\pm$ dictating the orientation of the Fermi rays, and $M_2(\theta_\pm) =  (\bm{\tau} \cdot \bm{e_{+}}) \sigma_+ + (\bm{\tau} \cdot \bm{e_{-}}) \sigma_{-}$, which dictates the curvature $\alpha = (\pi+\theta_+-\theta_-)/2$ of the Fermi arc, and controls the dispersion relation in the presence of a gap via the parameter $\theta_\tau = (\theta_+ + \theta_-)/2$. But some linear combinations of $M_1$ and $M_2$ also make suitable boundary conditions~\cite{faraei_greens_2018}. We find it necessary and sufficient for boundary matrices to be of the form
\begin{equation}
    M\left(\varphi,\theta^{(1)}_\pm,\theta^{(2)}_\pm\right) = M_1\left(\theta^{(1)}_\pm\right) \cos\varphi + M_2\left(\theta^{(2)}_\pm\right)\sin\varphi,
\end{equation}
with the constraint $\theta^{(1)}_+ - \theta^{(1)}_- -(\theta^{(2)}_+ - \theta^{(2)}_-) = \pi$
in order to satisfy $M^2 = M M^\dagger=\tau_0 \sigma_0$ and $\{M,\tau_z\sigma_z\}=0$. The additional parameter $\varphi$ controls the mixing between $M_1$ and $M_2$. We show in Fig.~\ref{figS:spectral_density} the surface spectral density maps at the Fermi level for a mixing with equal weight ($\varphi=\pi/4$) on both types $M_1$ and $M_2$. The surface still hosts Fermi arcs, though their shapes are slightly distorted. It also exhibits another branch of surface-localized modes with infinite extension, reminiscent of the Fermi rays that arise when pseudospin is utterly mixed under reflection.  

\section{Derivation of the local self-consistent Born approximation}

Let us now introduce the disorder-averaged Green's function $G$ which fulfils the boundary condition $MG(\epsilon,0,z')=G(\epsilon,0,z')$ in presence of impurities. We define the corresponding self-energy  operator $\Sigma$ through the equation 
\begin{equation}\label{eqS:def_G} 
[H_0 - \epsilon \tau_0 \sigma_0  +\Sigma(\epsilon,z,z')]G(\epsilon,z,z') = \delta(z-z') \tau_0 \sigma_0.
\end{equation}
In the SCBA, the bulk self-energy is momentum-independent, which entails in a semi-infinite system that $\Sigma$ depends on only one distance to the surface, e.g. $z$. In addition, we assume the self-energy to share the same chiral and pseudospin structure as the disorder potential itself, so that $\Sigma(\epsilon,z,z')= \Sigma(\epsilon,z) \tau_0 \sigma_0$ is proportional to the unit matrix.
We can perform calculations similar to that in Sec.~\ref{sec:Green_function} and  arrive at the following form for the off-diagonal components of the Green's function in the pseudospin sector :
\begin{equation}\label{eqS:Green_function} 
G^{\bar{\sigma} \sigma}_{\chi \chi'}(\epsilon,z,z') = \dfrac{\chi(k_x^\chi + i\sigma k_y^\chi)}{2(h'_{\chi \sigma}(z')+i\chi b_z)} \left[ \delta_{\chi \chi'} e^{-|h_{\chi \sigma}(z)-h_{\chi \sigma}(z')| - i\chi b_z|z-z'|} - A^{\bar{\sigma} \sigma}_{\chi \chi'}(z') e^{-[h_{\chi \sigma}(z) + h_{\chi \sigma}(z') + i\chi b_z(z+z')]}  \right],
\end{equation}
while the diagonal components read
\begin{equation}\label{eq:Green_ondiagnonal}
G^{\sigma \sigma}_{\chi \chi'}(\epsilon,z,z') = \dfrac{\epsilon-\Sigma(\epsilon, z) - i\chi\sigma \partial_z + \sigma b_z}{\chi(k_x^\chi + i\sigma k_y^\chi)} G^{\bar{\sigma} \sigma}_{\chi \chi'}(\epsilon,z,z').
\end{equation}
The function $A^{\bar{\sigma} \sigma}_{\chi \chi'}$ depends on the boundary condition.
For the $M_1$ boundary condition, only the diagonal chiral components are nonzero :
\begin{equation}\label{eqS:AM1} 
\left[A_{\chi \chi'}^{\bar{\sigma} \sigma}\right]_1(z') =  \dfrac{\epsilon-\Sigma_s(\epsilon)-i\chi\sigma h'_{\chi \sigma}(z') + 2\sigma b_z -\chi e^{-i\sigma\theta_\chi}(k_x^\chi + i\sigma k_y^\chi)}{\epsilon-\Sigma_s(\epsilon)+i\chi\sigma h'_{\chi \sigma}(z')-\chi e^{-i\sigma\theta_\chi}(k_x^\chi + i\sigma k_y^\chi)} \delta_{\chi \chi'},
\end{equation}
where $\theta_{\pm}$ are the two angles that parametrize $M_1$, and $\Sigma_s(\epsilon)=\Sigma(\epsilon,0)$ is the surface self-energy. For the $M_2$ boundary condition, the chiral components are coupled, and $A_{\chi \chi'}^{\bar{\sigma} \sigma}(z')$  takes a more complex form,
\begin{equation}\label{eqS:AM2} 
\left[A_{\chi \chi'}^{\bar{\sigma} \sigma} \right]_2(z') =   1- \dfrac{ 2i\chi \sigma e^{i\chi \theta_{\bar{\sigma}}} h'_{\chi \sigma}(z')(k_x^{\bar{\chi}} + i\sigma k_y^{\bar{\chi}}) }{D_{\chi \chi'}^{\bar{\sigma} \sigma} },
\end{equation}
where the denominator reads
\begin{equation}\label{eqS:Dpole} 
D_{\chi \chi'}^{\bar{\sigma} \sigma} = e^{i\chi \theta_{\bar{\sigma}}} \left(\epsilon - \Sigma_s(\epsilon) + i\chi \sigma h'_{\chi \sigma}(z')\right)(k_x^{\bar{\chi}} + i\sigma k_y^{\bar{\chi}}) + e^{i\chi \theta_{\sigma}} \left(\epsilon - \Sigma_s(\epsilon) + i\bar{\chi} \sigma h'_{\bar{\chi} \sigma}(z'))(k_x^{\chi} + i\sigma k_y^{\chi}\right).
\end{equation}
The function $h_{\chi \sigma}$ implicitly depends on the energy, and does not admit any simple analytical expression, but satisfies the differential equation 
\begin{equation}\label{eqS:h} 
({\bm{k}^\chi})^2 -(\epsilon-\Sigma)^2 + i\chi\sigma\Sigma' + h_{\chi \sigma}'' - (h_{\chi \sigma}')^2 = 0,
\end{equation}
with the boundary condition $h_{\chi \sigma}(0)=0$, and where $\Sigma'= \partial \Sigma/\partial z$. For a zero self-energy (in a clean system), Eq.~\eqref{eqS:h} can be solved and leads to $h_{\chi \sigma}(z)= q_\chi z$, in agreement with~Eq.~\eqref{eqS:G0_off}. In presence of disorder, Eq.~\eqref{eqS:h} cannot be solved to express analytically the function $h_{\chi \sigma}$ in terms of $\Sigma$ and its first derivatives. However, in the SCBA, the self-energy satisfies
\begin{equation}\label{eq:SCBA-app} 
\Sigma(\epsilon, z) = -\dfrac{\gamma }{4} \int_{k<\Lambda} \dfrac{\d^2 k}{(2\pi)^2} \text{Tr} \left[G(\epsilon,z,z)\right],
\end{equation}
 The SCBA relates the self-energy to a momentum integral of the trace of the Green's function, which depends itself implicitly on the whole function $z \mapsto \Sigma(\epsilon,z)$ : this problem is very hard to solve, even numerically. The scheme we refer to as the local self-consistent Born approximation (LSCBA) rests upon the following observation: assuming $\Sigma$ constant only in the integrand of the self-consistent equation, the Eq.~\eqref{eqS:Green_function} and~\eqref{eq:Green_ondiagnonal} reduce to the clean Green's function, up to the substitution $\epsilon \xrightarrow{} \epsilon-\Sigma$, i.e. $G(\epsilon,z,z') = G_0(\epsilon-\Sigma(\epsilon,z),z,z')$. The local quantity $\Sigma(\epsilon,z)$ is then related through the self-consistent equation to the function $\Sigma$ only at the same distance $z$ to the surface,
\begin{equation}\label{eq:LSCBA-app} 
\Sigma(\epsilon, z) = -\dfrac{\gamma }{4} \int_{k<\Lambda} \dfrac{\d^2 k}{(2\pi)^2} \text{Tr} \left[G_0(\epsilon-\Sigma(\epsilon,z),z,z)\right].
\end{equation}
Within this approximation, the SCBA is amenable to numerical solving.

\setcounter{equation}{0}
\section{Solution to the local self-consistent Born approximation}
\label{sec:LSCBAsolution}
While Eq.~(\ref{eq:LSCBA-app}) can be solved numerically we can further simplify it in the case $b=0$. To avoid any ambiguity we explicitly indicate the $(k,\phi)$ dependence of the Green's function, in contrast to our previous convention. The integrand of Eq.~\eqref{eq:LSCBA-app} naturally splits into two parts :
\begin{align}
\dfrac{\Sigma(\epsilon,z)}{\Delta} &= -\dfrac{\pi}{ \Lambda} \int_{k<\Lambda} \dfrac{\d^2k}{(2\pi)^2} \text{Tr}\left[ G_0(\epsilon-\Sigma(\epsilon,z),z,z,k,\phi) \right] \nonumber \\
&= -\dfrac{1}{4\pi\Lambda} \int_0^{\Lambda} k\d k \d \phi\,\text{Tr}  \left[G_{\rm b}(\epsilon-\Sigma(\epsilon,z),k,\phi) + \exp\left[{-2z\sqrt{k^2-(\epsilon-\Sigma(\epsilon,z))^2}}\right] G_{\rm e}(\epsilon-\Sigma(\epsilon,z),k,\phi) \right], \label{eqS:SCBA_split_be} 
\end{align}
where $G_{\rm b}(\epsilon,k,\phi) = [k(\tau_z\sigma_x \cos \phi + \tau_z\sigma_y \sin \phi) + \epsilon \sigma_0 \tau_0]/2\sqrt{k^2-\epsilon^2}$ is the bulk propagator of the free theory, and $G_{\rm e}(\epsilon,k,\phi)$ the excess contribution, which is a particular solution that satisfies the boundary condition. 
The first term of Eq.~\eqref{eqS:SCBA_split_be} is common to both boundary conditions, and can be fully integrated as
\begin{equation}
-\dfrac{1}{4\pi\Lambda} \int_0^{\Lambda} k\d k \d \phi\,\text{Tr}[G_{\rm b}(\epsilon-\Sigma,k,\phi)] = -\dfrac{1}{\Lambda} \int_0^{\Lambda}\dfrac{k\d k (\epsilon-\Sigma)}{\sqrt{k^2-(\epsilon-\Sigma)^2}} = \dfrac{\epsilon-\Sigma}{\Lambda}\left(\sqrt{-(\epsilon-\Sigma)^2}-\sqrt{\Lambda^2-(\epsilon-\Sigma)^2} \right).
\end{equation}
At the energy $\epsilon_{\rm F}(z)$ of the minimum of the local density, we have $\epsilon-\Sigma = i\Gamma(z)$ and the LSBA reduces to $\bar{\rho}(z) = s/\Delta = g(s,u) =  s\left(\sqrt{1+s^2}-s \right) + f(s,u)$ where we introduced the dimensionless variables $\Gamma(z) = \Lambda s$ and $z=u/\Lambda$, and the function $f(s,u)$ depends on the boundary condition. We will now express the contribution to the LSCBA from the excess propagator.

\paragraph{$M_1$ boundary condition}
For the $M_1$ boundary condition parametrized by the angles $\theta_+$ and $\theta_-$, the excess propagator $G_{\rm e}$ is diagonal in the chiral sector. Thus the trace $\text{Tr}[G_{\rm e}]$ splits into two partial traces over the chiral sectors, and for each partial trace, the integration variable $\bm{k}$ can be rotated by an arbitrary angle in order to absorb the dependence in $\theta_{\pm}$. As a result, we choose beforehand specific values of these angles to simplify the future computations, namely $\theta_+=0$ and $\theta_-=\pi$. Denoting generically by $S_{\rm e}$ the excess contribution to the LSCBA, we have
\begin{align}
S_{\rm e} &= -\dfrac{1}{4\pi\Lambda} \int_0^{\Lambda} k\d k \d \phi\,e^{-2z\sqrt{k^2-(\epsilon-\Sigma)^2}} \text{Tr}[G_{\rm e}(\epsilon-\Sigma,k,\phi)] \nonumber \\ &=  \dfrac{1}{2\pi\Lambda} \int_{0}^{\Lambda} k\d k\, e^{-2z\sqrt{k^2-(\epsilon-\Sigma)^2}} \int_{0}^{2\pi} \d \phi\, \left(\dfrac{\epsilon-\Sigma}{\sqrt{k^2-(\epsilon-\Sigma)^2}} + \dfrac{\sqrt{k^2-(\epsilon-\Sigma)^2}}{\epsilon-\Sigma-k \cos(\phi)}\right) \nonumber \\ 
&= \frac{e^{-2 z \sqrt{\Lambda^2-(\epsilon -\Sigma )^2}} \left( 2 i z \sqrt{\Lambda^2-(\epsilon -\Sigma )^2} + i -  2 z (\epsilon -\Sigma) \right) - e^{2 z \sqrt{-(\epsilon -\Sigma)^2}} \left(  2 i z \sqrt{-(\epsilon -\Sigma )^2}
+ i -  2 z (\epsilon -\Sigma) \right)  }{4 \Lambda z^2}.
\end{align}
Placing ourselves at the energy $\epsilon_{\rm F}(z)$ such that $\epsilon_{\rm F}(z)-\Sigma(\epsilon_{\rm F}(z),z)=i\Gamma(z)$ is imaginary, we find that $\epsilon_{\rm F}(z)-i\Lambda s = -i\Lambda \Delta \left[ s\left(\sqrt{1+s^2}-s \right) + f_1(s,u) \right]$ where
\begin{equation}
f_1(s,u)
=  \dfrac{1}{4 u^2} \left[e^{-2us}-e^{-2 u \sqrt{1+s^2} } \left(1+2u\left(\sqrt{1+s^2}-s\right)\right)\right].
\end{equation}
In particular, $\epsilon_{\rm F}(z)=0$ so that in Fermi rays the minimum of the local density is located at the Fermi level of the bulk nodes.
\\

\paragraph{$M_2$ boundary condition}
The $M_2$ boundary condition depends only on the parameter $\alpha=(\pi+\theta_+-\theta_-)/2$.
\begin{align}
S_{\rm e} &= \dfrac{1}{\Lambda} \int_{0}^{\Lambda} k\d k\, e^{-2z\sqrt{k^2-(\epsilon-\Sigma)^2}} \left(\dfrac{\epsilon-\Sigma}{\sqrt{k^2-(\epsilon-\Sigma)^2}} + \dfrac{\sqrt{k^2-(\epsilon-\Sigma)^2}}{\epsilon-\Sigma-k \cos(\alpha)}\right) \nonumber \\ 
&= \frac{(\epsilon-\Sigma)\left(e^{-2 z \sqrt{-(\epsilon -\Sigma )^2}}-e^{-2 z \sqrt{\Lambda^2-(\epsilon -\Sigma )^2}}\right) }{2\Lambda z} - \dfrac{1}{2\Lambda} \dfrac{\partial}{\partial z} \int_{0}^{\Lambda} k\d k\,   \dfrac{e^{-2z\sqrt{k^2-(\epsilon-\Sigma)^2}}}{\epsilon-\Sigma-k \cos(\alpha)} \nonumber .
\end{align}
The analytical computation can be pushed further if we restrict to the surface properties, e.g. to determine the surface density of Dirac states in Fig.~\ref{fig:FADoS}. Setting $z=0$, we find
\begin{align}
S_{\rm e} &=  \dfrac{\epsilon-\Sigma}{\Lambda}\left(\sqrt{\Lambda^2-(\epsilon-\Sigma)^2} - \sqrt{-(\epsilon-\Sigma)^2} \right) - \Lambda \sec(\alpha) - \sec(\alpha)^2 (\epsilon-\Sigma) \log\left(1-\dfrac{\Lambda \cos(\alpha)}{\epsilon-\Sigma}\right).
\end{align}
Placing ourselves at the energy $\epsilon_{\rm F}(z)$, we find that $\epsilon_{\rm F}(z)-i\Lambda s = -i\Lambda \Delta \left[ s\left(\sqrt{1+s^2}-s \right) + f_2(s,0) \right]$ where
\begin{multline}
f_2(s,0)
= s  \tan(\alpha)^2 \left(\sqrt{1+s^2}-s \right) + s^2 \sec(\alpha)^2\tan(\alpha)\left(\arctan(\cot(\alpha))-\arctan\left(\dfrac{s\tan(\alpha)}{\sqrt{1+s^2}}\right)\right) \\ - i \dfrac{\tan(\alpha)}{2} \left( 1-s^2 \sec(\alpha)^2 \log\left(1+\dfrac{\cos(\alpha)^2}{s^2}\right)\right).
\end{multline}
Hence the self-consistent equation reads $s/\Delta = s  \sec(\alpha)^2 (\sqrt{1+s^2}-s ) + s^2 \sec(\alpha)^2\tan(\alpha)(\arctan(\cot(\alpha))-\arctan(s\tan(\alpha)/\sqrt{1+s^2}))$, and enables to plot the surface density $\bar{\rho}(z)=s/\Delta $. The energy of the density minimum can then be found using $\epsilon_{\rm F}(z)=\tan(\alpha)( 1-s^2 \sec(\alpha)^2 \log(1+\cos(\alpha)^2/s^2))/2$.
\\

\setcounter{equation}{0}
\section{Computation of the group velocity}
One important property of the surface eigenstates is their group velocity $\bm{v}$. In this section, we show how to compute the Fermi velocity in a disordered sample, using the pole of the Green's function. Let us denote by $D(\epsilon, \bm{k})$ the pole of the excess Green's function. The pole depends on the boundary condition and is written explicitly in Eq.~\eqref{eqS:D1} for the $M_1$ BC and in Eq.~\eqref{eqS:D2} for the $M_2$ BC. Let $\epsilon_{\rm s}(\bm{k})$ be the dispersion relation of the surface eigenstates. By definition the group velocity of the surface energy band is $\bm{v} = \partial_{\bm{k}} \epsilon_s$~\cite{wilson_surface_2018}. Differentiating the equation $D(\epsilon_{\rm s},\bm{k})=0$, we relate the infinitesimal variations of energy and wavevector along the surface band structure, using $\d D = \partial_\epsilon D \, \d \epsilon + \partial_{\bm{k}} D  \cdot \d \bm{k} = 0$, which leads to $\bm{v} = -\partial_{\bm{k}} D/ \partial_\epsilon D$. In a dirty sample, the quasiparticles acquire a nonzero self-energy $\Sigma$, whose imaginary part determines the scattering rate and density of states, and real part the tweak of the relation dispersion induced by disorder. Within the LSCBA, the dispersion relation in presence of impurities reads $D(\epsilon_{\rm s}-\text{Re} \Sigma(\epsilon_{\rm s},0),\bm{k})=0$. In addition, the surface self-energy depends on the energy $\epsilon$ but not on the wavevector. Placing ourselves at the surface Fermi level $\epsilon_{\rm F}$ such that $\text{Re}\Sigma(\epsilon_{\rm F},0)=\epsilon_{\rm F}$, we find
\begin{subequations}
\begin{empheq}[left={\empheqlbrace\,}]{align}
\partial_\epsilon [D(\epsilon-\text{Re} \Sigma,\bm{k})](\epsilon_{\rm F}) &= (\partial_\epsilon D)(0,\bm{k})(1 - \partial_\epsilon (\text{Re}\Sigma)(\epsilon_{\rm F},0)),\\
\partial_{\bm{k}} [D(\epsilon-\text{Re} \Sigma,\bm{k})](\epsilon_{\rm F})  &= (\partial_{\bm{k}} D)(0,\bm{k}).
\end{empheq}
\end{subequations}
We finally express the group velocity $\bm{v}$ at the Fermi level in presence of disorder in terms of the same velocity $\bm{v_0}= -(\partial_{\bm{k}} D/ \partial_\epsilon D)(0,\bm{k})$ for the clean material : 
\begin{equation}
\bm{v} = \bm{v_0}[1 - \partial_\epsilon (\text{Re}\Sigma)(\epsilon_{\rm F},0)]^{-1}.
\end{equation}
This proves Eq.~\eqref{eq:v} in the main text. We now determine $\bm{v_0}$ for the Fermi rays, Fermi arcs, and surface Dirac states, using the relation dispersions found in Sec.~\ref{secS:M1BC} and Sec.~\ref{secS:M2BC}. 
Notice that since $\partial_{\bm{k}}D$ is orthogonal to any line of constant $D$, the group velocity is orthogonal to any slice of the surface band structure at fixed energy. We conclude that $\bm{v_0}$ is locally orthogonal to the Fermi rays, arc, or cone ; the remaining work consists in finding its norm.

\begin{figure}[b!]
\subfigure[]{\label{figS:vgrays}\includegraphics[scale=1]{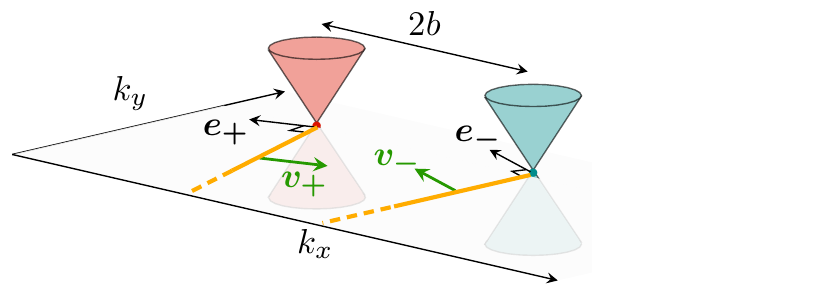}}\hfill
\subfigure[]{\label{figS:vgarc}\includegraphics[scale=1]{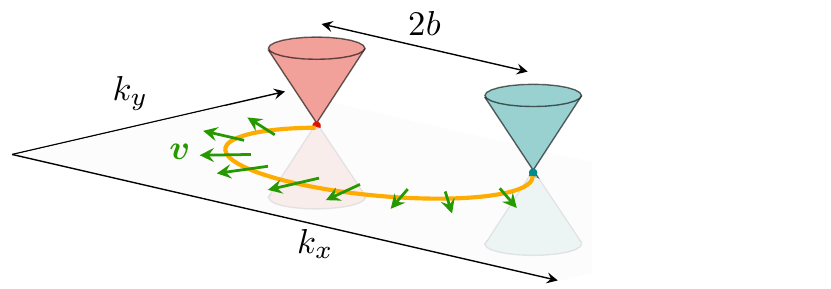}}
\caption{Schematic drawing of the group velocity of the surface-localized eigenstates. The states always propagate in a direction locally orthogonal to the surface band structure at the Fermi level. (a) Eigenstates populating a given Fermi ray all share the same group velocity and propagate in the same direction. (b) The eigenstates populating Fermi arcs propagate in various directions and with different velocities. The group velocity is minimal at the nodes and coincides with the Fermi velocity $v_0 = \cos \alpha$ of Dirac surface states.}
\end{figure}

\paragraph{Fermi rays} 
The group velocity for clean Fermi rays is $\bm{v_0^\chi} = -\partial_{\bm{k}} D_1/ \partial_\epsilon D_1$ where $D_1$ is given in Eq.~\eqref{eqS:D1} and $\chi=\pm 1$ labels the two rays. Plugging the relation dispersion \eqref{eqS:epsilon1}, we find $\bm{v_0^\chi} = - \chi \bm{e_\chi}$. The group velocity is depicted in Fig.~\ref{figS:vgrays}. Its norm is unity: the plane waves confined to the surface propagate with the same velocity as the bulk eigenstates.

\paragraph{Fermi arc} 
The group velocity for a clean Fermi arc is $\bm{v_0} = -\partial_{\bm{k}} D_2/ \partial_\epsilon D_2$. As for the Fermi rays, the group velocity is locally orthogonal to the Fermi arc, and points outward, as shown in Fig.~\ref{figS:vgarc}. Its norm is given by
\begin{equation}
v_0 = \dfrac{\sqrt{2} \cos(\alpha)}{\sqrt{1+\sqrt{1-(k_x \sin(2\alpha)/b)^2}}}.
\end{equation} 
The group velocity is maximal at the middle of the arc (for $k_x=0$), $v_{\rm max} = \sqrt{2} \cos(\alpha)/\sqrt{1+|\cos(2\alpha)|}$, and minimal at the nodes, $v_{\rm min}=\cos(\alpha)$.

\paragraph{Dirac Surface states} Since the relation dispersion of surface Dirac states is relativistic, the group velocity equates to the Fermi velocity $v_0 = \cos \alpha $.

\end{document}